\documentclass[aps,prb,reprint, superscriptaddress, showpacs]{revtex4-1}
\usepackage{amsmath}
\usepackage{amsfonts}
\usepackage{amssymb}

\begin{document}

\title{Non-equilibrium itinerant-electron magnetism: a time-dependent mean-field theory}

\author{A. Secchi}
\email{a.secchi@science.ru.nl , andrea.secchi@gmail.com}
\affiliation{Institute for Molecules and Materials, Radboud University Nijmegen, 6525 AJ Nijmegen, The Netherlands}

\author{A. I. Lichtenstein}
\affiliation{Institut f\"ur Theoretische Physik, Universitat Hamburg, Jungiusstra{\ss}e 9, D-20355 Hamburg, Germany}

\author{M. I. Katsnelson}
\affiliation{Institute for Molecules and Materials, Radboud University Nijmegen, 6525 AJ Nijmegen, The Netherlands}

\date{\today}

\begin{abstract}
We study the dynamical magnetic susceptibility of a strongly correlated electronic system in the presence of a time-dependent hopping field, deriving a generalized Bethe-Salpeter equation which is valid also out of equilibrium. Focusing on the single-orbital Hubbard model within the time-dependent Hartree-Fock approximation, we solve the equation in the non-equilibrium adiabatic regime, obtaining a closed expression for the transverse magnetic susceptibility. From this, we provide a rigorous definition of non-equilibrium (time-dependent) magnon frequencies and exchange parameters, expressed in terms of non-equilibrium single-electron Green functions and self-energies. In the particular case of equilibrium, we recover previously known results.
\end{abstract}


\maketitle

The dynamical magnetic susceptibility of an electronic system is a key quantity in both theoretical and experimental studies of magnetism \cite{Yosida, White}. In addition to its physical meaning as the first-order response function of the local magnetic moments to the application of a (space- and time-dependent) magnetic field, its relevance is due to the fact that its frequency spectrum contains all the magnetic excitations of the system. In particular, the spectrum of the transverse component of the magnetic susceptibility tensor contains the magnon frequencies.

To enable theoretical analysis, it is desirable to compute the magnon spectrum directly from a closed formula, rather than doing a numerical search of the poles of the transverse susceptibility. For strongly correlated systems in equilibrium, methods were developed to map electronic Hamiltonians onto effective classical spin models, from which one extracts the magnetic parameters (e.g. exchange) proper of the initial electronic systems when the magnetic moments undergo small rotations from their initial configuration \cite{Lichtenstein84, Lichtenstein87}; in the modern formulation, parameters are expressed in terms of single-electron Green functions (1EGFs) and self-energies \cite{Katsnelson00, Katsnelson02}. The original methods were recently extended to include unquenched electronic orbital degrees of freedom and relativistic interactions \cite{Katsnelson10, Secchi15, Secchi16, Udvardi03, Szilva13}. However, a direct connection between the magnetic parameters so determined and the poles of the transverse susceptibility is not obvious; within the framework of spin density functional theory, it has been shown that the original formulas yield accurate low-wavelength magnon frequencies for ferromagnetic systems, within the local spin-density approximation \cite{Katsnelson04}; corrections are required to compute thermodynamic properties \cite{Bruno03}.

Experimental progresses allow to modify the magnetic properties of materials by applying time-dependent fields coupling with the electrons, thereby modulating the magnetic interactions in time. Particularly, sub-picosecond laser fields \cite{Beaurepaire96, Hohlfeld97, Scholl97, Gudde99, Zhang00, Kimel05, Munzenberg10, Johnson12, Kirilyuk10} promise to provide the fastest possible modifications of magnetic states and, in the future, the fastest memory devices. Understanding how the magnetic properties are modulated in time requires a non-equilibrium microscopic theory of magnetism. Computationally, strongly correlated systems are typically treated with Dynamical Mean-Field Theory \cite{Metzner89, Georges92, Georges96, Kotliar06} or cluster perturbation theory \cite{Senechal00, Senechal02, Gros93} and their non-equilibrium formulations \cite{Freericks06, Aoki14, Eckstein10, MBalzer11}. At the moment, the computation of full non-equilibrium two-electron Green functions (2EGFs), such as the dynamical magnetic susceptibility, is not feasible due to huge memory requirements (even the computation of non-equilibrium 1EGFs is, in general, very demanding \cite{Balzer14}). To avoid the computation of 2EGFs, the mapping to a dynamical classical spin model has been proposed \cite{Secchi13}, where the time-dependent magnetic parameters are expressed in terms of non-equilibrium 1EGFs and self-energies. Also in this case, the connection to the magnetic susceptibility is not obvious.

In this Article we derive the self-consistent equation for the non-equilibrium magnetic susceptibility and solve it for the Hubbard model within the time-dependent Hartree-Fock approximation, in the adiabatic regime. We show that the effect of an external time-dependent field acting on the electrons (such as that of a laser or a phonon distribution) can be described by endowing the transverse magnetic susceptibility with time-dependent poles, i.e., time-dependent magnon frequencies. 

The remainder of this Article is organized as follows. In Section \ref{sec: notation} we introduce our notation and discuss the features of our non-equilibrium theory. We then present the problem in its most general formulation, before successively applying several approximations to reduce it to a solvable one. Therefore, in Section \ref{sec: GBSE} we introduce a generalized Bethe-Salpeter equation for the magnetic susceptibility, valid for arbitrary electronic models. The first two steps of approximations are taken in Section \ref{sec: THF}, where we apply the time-dependent Hartree-Fock approximation, and in Section \ref{sec: Hub}, where we restrict our theory to the (non-equilibrium) single-band Hubbard model. At this point, the problem can be solved in closed form in equilibrium, but not in the most general non-equilibrium case. The minimal non-equilibrium situation, which allows for a closed solution of the Bethe-Salpeter equation, corresponds to the adiabatic regime, which we introduce in Section \ref{sec: AD}. In this regime the system sustains time-dependent magnon excitations, meaning that the magnon frequencies are modulated in time by the action of the external field, but the magnon concept is still valid. In Section \ref{sec: exch} we characterize the non-equilibrium magnon frequencies by introducing non-equilibrium exchange parameters, and we recover well-known expressions valid in equilibrium as a particular case. In Section \ref{sec: Goldstone} we show that our theory is consistent with the Goldstone theorem even out of equilibrium. Finally, in Section \ref{sec: summary} we summarize our results and mention possible future extensions. In the Appendices we include the most technical passages of the derivations, which can be useful to the reader in order to reproduce our main results, but are not essential to follow the discussion in the main text.

\section{Notation} 
\label{sec: notation}

We formulate our non-equilibrium theory using the Kostantinov-Perel' (KP) time contour $\gamma = \gamma_+ \cup \gamma_- \cup \gamma_{\mathbb{M}}$, where $\gamma_{\pm}$ is the forward (backward) branch of the real-time (Keldysh) contour $\gamma_{\mathbb{K}} = \gamma_+ \cup \gamma_-$, and $\gamma_{\mathbb{M}}$ is the imaginary-time (Matsubara) branch \cite{KostantinovPerel, KadanoffBaym, Stefanucci, Kamenev}. If $z$ denotes a contour time variable, we write $z = t_{(\pm)}$ if $z$ lies on $\gamma_{\pm}$, where $t \in [ t_0, \infty)$ denotes a physical time; $t_{(+)} < t_{(-)}$ on the contour. We use letter $M$ to denote the third component of spin of the electron fields; $i, j, k$ to denote the sets of the other quantum numbers. $S_i^{\alpha}$ is the $\alpha$ component of the vector of spin matrices for the $i$-th field, with dimensionality $\mathcal{S}_i$. The non-equilibrium Hamiltonian is
\begin{align}
\hat{H}(z)  = \sum_{1 2} \hat{\psi}^{\dagger}_1 T^{1}_{2}(z) \, \hat{\psi}^2   + \frac{1}{4} \sum_{1 2 3 4} \hat{\psi}^{\dagger}_1 \hat{\psi}^{\dagger}_2 V^{2, 1}_{3, 4} \hat{\psi}^3 \hat{\psi}^4 
\label{Hamiltonian notation}
\end{align} 
for $z \in \gamma_{\mathbb{K}}$, where $1 \equiv \left(i_1 , M_1  \right)$ is a complete set of electron field indices, the single-electron terms depends on the contour coordinate $z$, and the interaction matrix element is antisymmetrized, $V^{2, 1}_{3, 4} = V^{1, 2}_{4, 3} = - V^{1, 2}_{3, 4} = - V^{2, 1}_{4, 3}$. The single-electron Hamiltonian includes the time-dependent terms generated by the coupling of the electrons with an external time-dependent field. On the Matsubara branch, the Hamiltonian may have a different form \cite{Stefanucci}, which we denote, in general, as
\begin{align}
\hat{H}(z) = \hat{H}_{\mathbb{M}},
\end{align}
independent of $z$ for $z \in \gamma_{\mathbb{M}}$. The Hamiltonian on the Matsubara branch should be considered as a tool to prepare the system in some known state at the initial time $t_0$; it might coincide (up to conserved quantities) with the physical Hamiltonian at the initial time $t_0$, in which case the system is prepared in a thermal superposition. Alternatively, one can choose $\hat{H}_{\mathbb{M}}$ as an effective projector over a state or a set of states of interest. For example, to prepare the system in a fully spin-polarized state, one can include in $\hat{H}_{\mathbb{M}}$ a Zeeman term coupling the spins with an auxiliary uniform magnetic field, despite the fact that the Hamiltonian of the system of interest (on the real-time branches) might not include such magnetic field. Also taking a low temperature, this effectively restricts the system to a broken-symmetry configuration, which would not be captured in the absence of the auxiliary magnetic field. The results that we present in this work hold independently of the particular choice of the Hamiltonian on the Matsubara branch.

1EGFs and 2EGFs are denoted as
\begin{align}
 & G^{1 z_1}_{2 z_2} \equiv - \mathrm{i} \left< \mathcal{T}_{\gamma} \hat{\psi}^{1   z_1} \hat{\psi}^{\dagger}_{2   z_2}  \right>, \nonumber \\
 & G^{1 z_1, 3 z_3}_{2 z_2, 4 z_4} \equiv \left( - \mathrm{i} \right)^2  \left< \mathcal{T}_{\gamma} \hat {\psi}^{1 z_1} \hat{\psi}^{3 z_3} \hat{\psi}^{\dagger}_{4 z_4} \hat{\psi}^{\dagger}_{2 z_2}  \right>,
\end{align}
where $\left< \ldots \right>$ denotes an expectation value computed using the contour evolution operators \cite{KostantinovPerel, KadanoffBaym, Stefanucci}. The contour 1EGFs are related to the lesser/greater Green functions via
\begin{align}
&  G^{1 z_1}_{2 z_2} \equiv \Theta(z_1, z_2) \left( G^> \right)^{1 t_1}_{2 t_2} + \Theta(z_2, z_1) \left( G^< \right)^{1 t_1}_{2 t_2}, \nonumber \\
&  \left( G^> \right)^{1 t_1}_{2 t_2} = - \mathrm{i} \left< \hat{\psi}^{1 t_1} \hat{\psi}^{\dagger}_{2 t_2} \right>, \quad \left( G^< \right)^{1 t_1}_{2 t_2} =   \mathrm{i} \left< \hat{\psi}^{\dagger}_{2 t_2} \hat{\psi}^{1 t_1}  \right> ,
\label{G Keldysh}
\end{align} 
where $\Theta(z_1, z_2)$ is the step function on the KP contour. Finally, the Dyson equation reads
\begin{align}
& \mathrm{i} \partial_{z} G^{1 z}_{2 z_2}   -   \sum_3 \! \left[ T^{1 }_{3}(z)       G^{3 z}_{2 z_2}   +    \!  \int_{\gamma} \! \mathrm{d} z_3  \Sigma^{1 z}_{3 z_3 } G^{3 z_3 }_{2 z_2}  \right]
 \!  =   \delta^1_2 \delta(z, z_2)                , 
\label{eq motion GF}
\end{align}
where the self-energy $\Sigma$ is defined via
\begin{align}
\sum_5 \int_{\gamma} \mathrm{d} z_5 \Sigma^{1 z_1}_{5 z_5} G^{5 z_5}_{2 z_2} \equiv \frac{\mathrm{i}}{2}  \sum_{ 3 4 5  }   V^{3 , 1}_{4  , 5}  G^{4  z_1   , 5  z_1 } _{2 z_2, 3 ( z_1 + \epsilon )}      .
\end{align}

\section{Generalized Bethe-Salpeter equation}
\label{sec: GBSE}

The dynamical magnetic susceptibility tensor is  
\begin{align}
\chi^{\alpha \alpha'}_{i j}(t, t') & \equiv  \left. \frac{\delta \left< \hat{S}_i^{\alpha}(t) \right>_{\boldsymbol{B}}}{\delta B_j^{\alpha'}(t')} \right|_{\boldsymbol{B} = \boldsymbol{0}} \nonumber \\
& = - \mathrm{i} \Theta(t - t') \left< \left[ \hat{S}^{\alpha}_i(t), \hat{S}^{\alpha'}_j(t') \right] \right>,
\label{def chi}
\end{align}
where $\left< \ldots \right>_{\boldsymbol{B}}$ denotes an expectation value computed in the presence of the magnetic field $\boldsymbol{B} \equiv \lbrace \boldsymbol{B}_i(t) \rbrace$ coupling with the spins, and $\hat{S}_i^{\alpha}(t)$ is the $\alpha$ component of the $i$-th spin operator of the system at time $t$; $\alpha \in \lbrace x, y, z \rbrace$ or $\alpha \in \lbrace +, -, z \rbrace$. The second line of Eq.\eqref{def chi} is the Kubo formula, which connects $\chi^{\alpha \alpha'}_{i j}(t, t')$ to relevant many-body quantities. For example, for a ferromagnetic lattice in equilibrium, the low-energy poles of the Laplace transform of the transverse magnetic susceptibility $\chi^{+ -}_{\boldsymbol{q}}(\omega)$ are the magnon frequencies $\omega_{\boldsymbol{q}}$.

We now generalize the Bethe-Salpeter Equation (BSE) for the magnetic susceptibility to the case of the most arbitrary electronic system out of equilibrium. It is convenient to define the matrices
\begin{align}
& \chi^{\alpha \alpha'}_{1, 2 ; j }(z_1, z_2 ; z_3) \equiv - \mathrm{i}   \left. \frac{\delta    \left( S^{\alpha}  \cdot  G_{\boldsymbol{B}}\right)^{1 z_1}_{2 z_2} }{\delta B_{j z_3}^{\alpha'}} \right|_{\boldsymbol{B} = \boldsymbol{0}}     , \nonumber \\
& \chi^{\alpha \alpha'}_{1, 2 ; j }(z_1, z_2 ; t') \equiv \chi^{\alpha \alpha'}_{1, 2 ; j }(z_1, z_2 ; t'_{(+)}) - \chi^{\alpha \alpha'}_{1, 2 ; j }(z_1, z_2 ; t'_{(-)})  ,
\label{most general chi}
\end{align}
where the magnetic field is allowed to take different values for the two Keldysh coordinates corresponding to the same physical time. The susceptibility matrix defined in Eq.\eqref{most general chi} satisfies the following Generalized Bethe-Salpeter Equation (GBSE) on the KP contour [for the full derivation, see Appendix \ref{app: GBSE}],
\begin{align}
& \chi^{\alpha \alpha'}_{1, 2 ; j }(z_1, z_2 ; t') =  \left( \chi_0 \right)^{\alpha \alpha'}_{1, 2 ; j }(z_1, z_2 ; t') \nonumber \\ 
& + \sum_{\alpha''  \alpha'''}  \sum_{4 5 6 7}   \int_{\gamma} \mathrm{d} \left(w_4 ,   w_5  , w_6 , w_7 \right)\left( \chi_0^{\alpha \alpha''} \right)^{1 z_1, 5 w_5}_{2 z_2  , 4 w_4 }    \nonumber \\
& \times \left( \Gamma^{\alpha'' \alpha'''} \right)^{4 w_4, 7w_7}_{5 w_5, 6 w_6}    \chi^{\alpha''' \alpha'}_{6, 7 ; j}(w_6, w_7 ; t')  ,
\label{BS t'}
\end{align}
where we have introduced the quantities
\begin{align}
\left( \chi_0^{\alpha \alpha''} \right)^{1 z_1, 5 w_5}_{2 z_2  , 4 w_4  }  \equiv - \mathrm{i} \left(  S^{\alpha}  \cdot G  \cdot S^{\alpha''}  \right)^{1 z_1}_{4 w_4}   G^{5 w_5}_{2 z_2}  ,
\label{the chi_0}
\end{align}
\begin{align}
& \left( \Gamma^{\alpha'' \alpha'''} \right)^{4 w_4, 7w_7}_{5 w_5, 6 w_6}  \equiv  \frac{   \mathrm{i} }{\mathcal{S}_{i_4} \left( \mathcal{S}_{i_4} + 1 \right)   }  \,  \frac{\delta \left(    S^{\alpha''} \cdot  \Sigma \right)^{4 w_4}_{5 w_5}   }{\delta \left( S^{\alpha'''} \cdot G \right)^{6 w_6}_{7 w_7} }  ,
\label{the Gamma}
\end{align}
\begin{align}
& \left( \chi_0 \right)^{\alpha \alpha'}_{1, 2 ; j }(z_1, z_2 ; t' )     = \sum_{s = \pm} s  \sum_M  \left( \chi_0^{\alpha \alpha'} \right)^{1 z_1,  \,  j M t'_{(s)}}_{ 2 z_2 , \, j M t'_{(s)}   }   .
\label{main chi_0}
\end{align}
The physical susceptibility given by Eq.\eqref{def chi} can be obtained from Eqs.\eqref{most general chi} via the relation
\begin{align}
&  \chi^{\alpha \alpha'}_{i j}(t , t')     =   \sum_M   \chi^{\alpha \alpha'}_{i M, i M ; j }(t_{(+)}, t_{(-)} ; t' )  ,
\label{physical}
\end{align}
as detailed in Appendix \ref{app: dynsusc}.

\section{Time-dependent Hartree-Fock approximation}
\label{sec: THF}

Equation \eqref{BS t'} is exact, but its matrix structure is very complicated. The time-dependent Hartree-Fock approximation (THF) \cite{Stefanucci} greatly simplifies its time-domain structure. In THF, the 2EGF is approximated as
\begin{align}
  G^{A, C}_{B, D} \stackrel{\mathrm{THF}}{=} G^A_B G^C_D - G^A_D G^C_B , 
\label{THF definition}
\end{align}
which yields the following expression for the self-energy:
\begin{align}
 &    \Sigma^{1 z_1}_{2 z_2}   \stackrel{\mathrm{THF}}{=}  -  \mathrm{i} \delta(z_1, z_2)      \sum_{ 3 4    }      V^{1 , 3}_{2  , 4}    \left( G^< \right)^{  4  t_1 } _{  3 t_1  } . 
    \label{Sigma HF}
\end{align}
It should be noted that the 1EGFs appearing in Eqs.\eqref{THF definition} and \eqref{Sigma HF} are \emph{not} the non-interacting Green functions which are used in the conventional many-body perturbation theory for weakly correlated systems, where the electron-electron interaction is the small parameter. In that case, Eqs.\eqref{THF definition} and \eqref{Sigma HF} would reduce to the RPA scheme. In our case, instead, single-particle Green functions are the solutions of an \emph{interacting} problem, although simplified via the THF approximation. This amounts to the only approximation that the self-energy is local in time. Although the equations are formally similar, this difference between THF and RPA is crucial to properly describe the magnon excitations for strongly correlated systems.

With this distinction in mind, we now introduce
\begin{align}
\left( \chi_0 \right)^{\alpha \alpha'}_{i  j }(t , t')  =  \sum_{M} \left( \chi_0 \right)^{\alpha \alpha'}_{i M, i M ; j }(t_{(+)}, t_{(-)} ; t')  , 
\label{chi_0 physical}
\end{align}
which is a physical quantity defined in terms of 1EGFs, whose meaning depends on the approximation scheme.  In our case, it can be called \emph{Stoner susceptibility}, since its spectrum contains only electron-hole excitations which are analogous of those of the Stoner theory for the Hubbard model. In contrast, within RPA Eq.\eqref{chi_0 physical} would coincide with the bare magnetic susceptibility of a non-interacting system.

Applying Eq.\eqref{Sigma HF} to Eq.\eqref{the Gamma}, we obtain
\begin{align}
& \left( \Gamma^{\alpha'' \alpha'''}  \right)^{4 w_4, 7w_7}_{5 w_5, 6 w_6}  \nonumber \\
& \stackrel{\mathrm{THF}}{=}        \delta(w_4, w_5)  \delta(w_4, w_6) \delta(w_7, w_4 + \epsilon) \left( \Gamma_{\mathrm{THF}}^{\alpha'' \alpha'''} \right)^{4, 7}_{5, 6} ,
\end{align}
where
\begin{align}
\left( \Gamma_{\mathrm{THF}}^{\alpha'' \alpha'''} \right)^{4, 7}_{5, 6} \equiv  \sum_{ M  M'} \frac{  \left( S_{i_4}^{\alpha''}  \right)^{M_4}_{M'}      V^{(i_4 M') , 7}_{5  , (i_6 M)}      \left( S_{i_6}^{\alpha'''} \right)_{M_6}^M  }{\mathcal{S}_{i_4} \left( \mathcal{S}_{i_4} + 1 \right) \mathcal{S}_{i_6} \left( \mathcal{S}_{i_6} + 1 \right) }      .
\label{exchange field HF}
\end{align}
Inserting Eq.\eqref{exchange field HF} into Eq.\eqref{BS t'} yields the THF form of the GBSE,
\begin{align}
& \chi^{\alpha \alpha'}_{1, 2 ; j }(z_1, z_2 ; t' ) \stackrel{\mathrm{THF}}{=}  \left( \chi_0 \right)^{\alpha \alpha'}_{1, 2 ; j }(z_1, z_2 ; t') \nonumber \\ 
& + \sum_{\alpha''  \alpha'''}  \sum_{4 5 6 7}   \int_{\gamma} \mathrm{d}w_4  \left( \chi_0^{\alpha \alpha''} \right)^{1 z_1, 5 w_4}_{2 z_2  , 4 w_4 }      \left( \Gamma_{\mathrm{THF}}^{\alpha'' \alpha'''} \right)^{4, 7}_{5, 6}  \nonumber \\
& \times \chi^{\alpha''' \alpha'}_{6, 7 ; j }(w_4 , w_4 + \epsilon ; t')   .
\label{BSalp HF 1}
\end{align}
If $t_4$ is the physical time corresponding to the contour coordinate $w_4$, then the quantity
\begin{align}
\chi^{\alpha''' \alpha'}_{6, 7 ; j }(w_4 , w_4 + \epsilon ; t')  \equiv \chi^{\alpha''' \alpha'}_{6, 7 ; j }(t_4 ; t') 
\label{simpler chi} 
\end{align}
depends only on $t_4$, independently of the contour branch on which $w_4$ lies. This can be seen by applying Eq.\eqref{most general chi} with $z_1 = w_4$ and $z_2 = w_4 + \epsilon$. In terms of Eq.\eqref{simpler chi}, we have
\begin{align}
& \chi^{\alpha \alpha'}_{i  j }(t , t')  \equiv \sum_M  \chi^{\alpha \alpha'}_{i M, i M ; j }(t ; t' )        .
\end{align}
Putting $z_1 = t_{(+)}$ and $z_2 = t_{(-)}$, and 
\begin{align}
&  \chi^{\alpha \alpha'}_{1, 2 ; j }(t ; t' ) \equiv \chi^{\alpha \alpha'}_{1, 2 ; j }(t_{(+)}, t_{(-)} ; t' ) , \nonumber \\
&  \left( \chi_0 \right)^{\alpha \alpha'}_{1, 2 ; j }(t ; t') \equiv \left( \chi_0 \right)^{\alpha \alpha'}_{1, 2 ; j }(t_{(+)}, t_{(-)} ; t'),
\end{align}
we then obtain, from Eq.\eqref{BSalp HF 1}, the THF GBSE in real-time coordinates:
\begin{align}
& \chi^{\alpha \alpha'}_{1, 2 ; j }(t ; t' ) \stackrel{\mathrm{THF}}{=}  \left( \chi_0 \right)^{\alpha \alpha'}_{1, 2 ; j }(t ; t') \nonumber \\ 
& \! + \! \! \sum_{\alpha''  \alpha'''} \! \sum_{4 5 6 7}   \int_{t_0}^{\infty} \! \! \mathrm{d}t'' \chi_0^{\alpha \alpha''}\!(t, t'')^{1 , 5   }_{2    , 4   }   \left( \Gamma_{\mathrm{THF}}^{\alpha'' \alpha'''} \right)^{4, 7}_{5, 6}   \chi^{\alpha''' \alpha'}_{6, 7 ; j }\!(t'' ; t')   ,
\label{BSalp HF 2}
\end{align}
where we have converted the contour integration to physical-time integration, we have used the fact that $\chi^{\alpha''' \alpha'}_{6, 7 ; j }(w_4 , w_4 + \epsilon ; t') = 0$ if $w_4 \in \gamma_{\mathrm{M}}$, and we have introduced
\begin{align}
&  \chi_0^{\alpha \alpha''}\!(t, t'')^{1 , 5   }_{2    , 4   }   \equiv \left( \chi_0^{\alpha \alpha''} \right)^{1 t_{(+)}, 5 t''_{(+)}  }_{2 t_{(-)}  , 4 t''_{(+)} }          - \left( \chi_0^{\alpha \alpha''} \right)^{1 t_{(+)}, 5 t''_{(-)}}_{2 t_{(-)}  , 4 t''_{(-)} }  \nonumber \\
 & = - \mathrm{i} \theta(t - t'')    \Big[  \left(  S^{\alpha}  \cdot   \left( G^{>} \right)^t_{t''} \cdot S^{\alpha''}  \right)^{1  }_{4  }   \left( G^< \right)^{5 t'' }_{2 t }    \nonumber \\
 & \quad  - \left(  S^{\alpha}  \cdot   \left( G^< \right)^t_{t''}  \cdot S^{\alpha''}  \right)^{1  }_{4  }   \left( G^{>} \right)^{5 t'' }_{2 t } \Big]    ,
 \label{another one}
\end{align}
and therefore
\begin{align}
\left( \chi_0 \right)^{\alpha \alpha'}_{1, 2 ; j }(t ; t') = \sum_M  \chi_0^{\alpha \alpha'}\!(t, t')^{1 , j M   }_{2  , j M  } .
\label{bare susceptibility matrix definition}
\end{align}

\section{Single-orbital Hubbard model}
\label{sec: Hub}

To achieve a further simplification, we restrict our theory to the single-orbital Hubbard model (SOH). In this case the spin space has dimensionality $\mathcal{S} = 1/2$ at every site, and the interaction Hamiltonian becomes
\begin{align}
\hat{V} = \frac{1}{4} \sum_{3 4 5 6} V^{4, 3}_{5, 6} \hat{\psi}^{\dagger}_3  \hat{\psi}^{\dagger}_4 \hat{\psi}^5 \hat{\psi}^6 \stackrel{\mathrm{SOH}}{=} \sum_{i} U_i \hat{n}_{i \uparrow} \hat{n}_{i \downarrow} ,
\end{align}
which implies
\begin{align}
  V^{4, 3}_{5, 6} \stackrel{\mathrm{SOH}}{=}    \delta^{i_4}_{i_5}  \delta^{i_3}_{i_5} \delta^{i_3}_{i_6} 
 \,  \delta^{M_4}_{\overline{M_3}} \, \delta_{M_6}^{\overline{M_5}} \, \left( \delta_{M_5}^{\overline{M_3}} - \delta_{M_5}^{ M_3 }     \right)
  U_{i_3}  .
  \label{Hubbard V}
\end{align}
The SOH Hamiltonian is spin-independent: $\left[ \hat{H}(t), \hat{S}^z \right] = 0$, so the total third component of the spin of the system is a good quantum number.
The transverse component of Eq.\eqref{BSalp HF 2}, corresponding to $(\alpha, \alpha') = (+, -)$, then simplifies as   
\begin{align}
  & \chi^{+ -}_{i  j }(t , t' ) \stackrel{\mathrm{THF}}{=} \, \stackrel{\mathrm{SOH}}{=}  \left( \chi_0 \right)^{+-}_{i  j }(t , t') \nonumber \\
  &  +   \sum_{k}   \int_{t_0}^{\infty} \mathrm{d}t''  \left( \chi_0 \right)^{+ -}_{i k}(t, t'') \left( - U_k \right) \chi^{+ -}_{k  j }(t'' , t')   ,
\label{BSalp THF SOH}
\end{align}
which is the non-equilibrium SOH version of the equation used in Ref.\onlinecite{Katsnelson04}. Compared to the general case, this form of the BSE has the simplest possible structure in both time and spin domains. Details about the derivation of Eq.\eqref{BSalp THF SOH} are given in Appendix \ref{app: BSE Hub}.

The Stoner transverse susceptibility is
\begin{align}
\left( \chi_0 \right)^{+-}_{i  j }(t , t') \stackrel{\mathrm{SOH}}{=} & \, \mathrm{i} \theta(t - t') \Big[ \left( G^< \right)^{i \downarrow t}_{j \downarrow t'}    \left( G^> \right)_{i \uparrow t}^{j \uparrow t'}   \nonumber \\
&     - \left( G^> \right)^{i \downarrow t}_{j \downarrow t'}    \left( G^< \right)_{i \uparrow t}^{j \uparrow t'} \Big] .
\label{chi_0 SOH}
\end{align} 
We now derive an effective equation for this quantity, by applying the operator $- \mathrm{i} \partial_{t'}$ to Eq.\eqref{chi_0 SOH} and using the Dyson equations in the THF approximation, which read as
\begin{align}
&  - \mathrm{i} \partial_{t'} \! \left( G^{\lessgtr} \right)^{i M t}_{j M t'} \stackrel{\mathrm{THF}}{=} \! \left( G^{\lessgtr} \cdot T \right)^{i M t}_{j M t'}  + \! \left( G^{\lessgtr} \right)^{i M t}_{j M t'} \Sigma_{j M}(t') , \nonumber \\
&    \mathrm{i} \partial_{t'} \left( G^{\lessgtr} \right)_{i M t}^{j M t'} \stackrel{\mathrm{THF}}{=} \left( T \cdot G^{\lessgtr}  \right)_{i M t}^{j M t'}  + \Sigma_{j M}(t')  \left( G^{\lessgtr} \right)_{i M t}^{j M t'}  ,
\label{Dyson THF SOH}
\end{align} 
where $\Sigma_{j M}(t') \equiv U_j \rho_{j \overline{M}}(t')$ is the THF self-energy for the SOH model, with $\rho_{j \overline{M}}(t') \equiv \left< \hat{\psi}_{j \overline{M}}^{\dagger}(t') \hat{\psi}^{j \overline{M}}(t') \right>$. We then obtain
\begin{align}
& \left( \chi_0 \right)^{+ -}_{i  j}(t, t') \left[ - \mathrm{i} \overleftarrow{\partial_{t'}}  - \Delta_j(t') \right]  \nonumber \\
&   \stackrel{\mathrm{SOH}}{=} \, \stackrel{\mathrm{THF}}{=}  \delta(t - t') \delta_{i j} m_j(t') +  \Lambda_{i j}(t, t')      ,
\label{bare susceptibility i j t t'}
\end{align}
where $m_j(t') \equiv \rho_{j \uparrow}(t') - \rho_{j \downarrow}(t')$,  
\begin{align}
 \Delta_j(t') \equiv U_j m_j(t') \equiv - 2 \Sigma_{j S}(t') \equiv \Sigma_{j \downarrow}(t') - \Sigma_{j \uparrow}(t')  
 \label{Stoner}
\end{align}
is the time-dependent Stoner splitting, and 
\begin{widetext} 
\begin{equation}
\Lambda_{i j}(t, t') =  
 \mathrm{i} \theta(t - t')  \Big[  \!  \left( G^> \right)^{i \downarrow t}_{j \downarrow t'}        \!   \left( T \cdot G^< \right)_{i \uparrow t}^{j \uparrow t'}  
    -       \left( G^< \right)_{j \downarrow t'}^{i \downarrow t}     \!      \left( T \cdot G^> \right)^{j \uparrow t'}_{i \uparrow t} 
     -       \left( G^>\cdot T  \right)^{i \downarrow t}_{j \downarrow t'}  \!  \left( G^< \right)_{i \uparrow t}^{j \uparrow t'}    
    +    \left( G^< \cdot T \right)_{j \downarrow t'}^{i \downarrow t}  \!  \left( G^> \right)^{j \uparrow t'}_{i \uparrow t}       \Big]  .
\label{Lambda GF}             
\end{equation}
\end{widetext}
We now determine the transverse magnetic susceptibility from the BSE, Eq.\eqref{BSalp THF SOH}, and the approximate equation for the bare susceptibility, Eq.\eqref{bare susceptibility i j t t'}.

\section{Adiabatic approximation and non-equilibrium magnons}
\label{sec: AD}

We introduce the Wigner time coordinates, $\tau \equiv t - t'$ and $T \equiv \left( t + t' \right) / 2$, which are respectively called the relative time and the total time. In this section we send the initial time $t_0 \rightarrow - \infty$, so that the domain of $T$ is $\left( - \infty, \infty \right)$, and we can define the Fourier transforms with respect to $T$ on the whole real axis. We put $f(t, t') \equiv \widetilde{f}(\tau , T)$ to distinguish the representations of a function in terms of the individual fermionic time arguments versus the Wigner coordinates. We apply to both Eqs.\eqref{BSalp THF SOH} and \eqref{bare susceptibility i j t t'} the Laplace transform with respect to $\tau$ and the Fourier transform with respect to $T$. We use the notation
\begin{align}
\widetilde{f}(\omega , \Omega) \equiv \int_{-\infty}^{\infty}   \mathrm{d} T \, \mathrm{e}^{\mathrm{i} \Omega T} \int_0^{\infty} \mathrm{d} \tau \, \mathrm{e}^{\mathrm{i} \omega \tau} \widetilde{f}(\tau , T) ,
\end{align}
where $\Im\left(\omega\right) > 0$. We obtain the following representations of Eqs.\eqref{BSalp THF SOH} and \eqref{bare susceptibility i j t t'} in the frequency domain [the full derivation can be found in Appendix \ref{app: frequency derivation}]:
\begin{widetext}
\begin{align}
  & \widetilde{\chi}^{+ -}_{i   j }(\omega , \Omega ) - \left( \widetilde{\chi}_0 \right)^{+-}_{i   j }(\omega , \Omega) 
  \stackrel{\mathrm{THF}}{=} \, \stackrel{\mathrm{SOH}}{=}          -    
  \int_{-\infty}^{\infty} \frac{\mathrm{d} \Omega'}{2 \pi}        \sum_{k}  \left( \widetilde{\chi}_0 \right)^{+ -}_{i  k}\left(  \omega + \frac{\Omega - \Omega'}{2} ,  \Omega' \right)       \, U_k \, 
 \widetilde{\chi}^{+ -}_{k   j }\left( \omega    - \frac{\Omega'}{2} , \Omega - \Omega'  \right)  ,
  \label{Bethe-Salpeter i j omega Omega}
\end{align}
\begin{align}
&  \left(   \omega - \frac{\Omega}{2}     \right)     \left( \widetilde{\chi}_0 \right)^{+ -}_{i  j}(\omega , \Omega)  
 -      \int_{- \infty}^{\infty} \frac{\mathrm{d} \Omega'}{2 \pi}        \left( \widetilde{\chi}_0 \right)^{+ -}_{i  j}\!\left( \omega    +   \frac{\Omega - \Omega'}{2} ,  \Omega'  \right) \,  \Delta_j\!\left( \Omega - \Omega' \right)  
  \stackrel{\mathrm{THF}}{=}  \, \stackrel{\mathrm{SOH}}{=}   \delta^i_j m_j(\Omega) +  \widetilde{\Lambda}_{i j}( \omega , \Omega)   . 
\label{bare susceptibility i j omega Omega}      
\end{align}
\end{widetext}

In the non-equilibrium adiabatic (AD) regime, we assume that the susceptibilities are non-zero only when the frequencies related to the Fourier transforms with respect to the total time $T$ are much smaller than the frequencies related to the Laplace transforms with respect to the relative time $\tau$. In this case, Eq.\eqref{bare susceptibility i j omega Omega} simplifies into
\begin{align}
&          \left( \widetilde{\chi}_0 \right)^{+ -}_{i  j}(\omega ; T)        \stackrel{\mathrm{THF}}{=}  \,  \stackrel{\mathrm{SOH}}{=} \, \stackrel{\mathrm{AD}}{=}  
\frac{ \delta_{i j} m_j(T) +  \widetilde{\Lambda}_{i j}( \omega ; T )  }{ \omega - \Delta_j(T) }  ,
\label{bare susceptibility i j omega T adiabatic}      
\end{align}
while Eq.\eqref{Bethe-Salpeter i j omega Omega} simplifies into
\begin{align} 
  &      \sum_{k} \left[ \delta_{i  k} + \left( \widetilde{\chi}_0 \right)^{+ -}_{i  k}\left(  \omega   ; T \right)       \, U_k \right] \widetilde{\chi}^{+ -}_{k   j }\left( \omega ; T  \right)   \nonumber \\
  & \stackrel{\mathrm{THF}}{=} \, \stackrel{\mathrm{SOH}}{=}   \, \stackrel{\mathrm{AD}}{=}    \left( \widetilde{\chi}_0 \right)^{+-}_{i   j }(\omega ; T)   .
  \label{Bethe-Salpeter i j omega T adiabatic}
\end{align}
We substitute Eq.\eqref{bare susceptibility i j omega T adiabatic} into Eq.\eqref{Bethe-Salpeter i j omega T adiabatic} and, after some algebra, we get
\begin{align}
      \widetilde{\chi}^{+ -}_{i   j }\left( \omega  ; T     \right)   \stackrel{\mathrm{THF}}{=} \, \stackrel{\mathrm{SOH}}{=}  \, \stackrel{\mathrm{AD}}{=} \, & \frac{\omega - \Delta_i(T)}{\omega - \Delta_j(T)}  \sum_k \frac{U_k}{U_i}   \overrightarrow{F}^{-1}_{i k}(\omega ; T)  \nonumber \\
 & \times \left[   \delta_{k j} m_j(T)  +   \widetilde{\Lambda}_{k j}( \omega  ; T  )   \right] ,
  \label{Bethe-Salpeter i j omega adiabatic}
\end{align}
where we have introduced the matrix
\begin{align}
 F_{i  k}(\omega ; T ) \equiv      \delta_{i  k} \omega +  U_i \widetilde{\Lambda}_{i k}( \omega ; T )        
\label{f matrix adiabatic}
\end{align}
and its left inverse $\overrightarrow{F}^{-1}(\omega ; T)$, defined via
\begin{align}
\sum_{i} \overrightarrow{F}^{-1}_{l  i}(\omega ; T)  \, F_{i  k}(\omega ; T) = \delta_{l k} .
\end{align}

The susceptibility has a pole when the matrix \eqref{f matrix adiabatic} has a null eigenvalue. If we assume that $\widetilde{\Lambda}_{i k}(\omega ; T)$ is almost independent of $\omega$ at frequencies much smaller than the Stoner excitations, then the poles are obtained when $\omega$ is an eigenvalue of the time-dependent matrix
\begin{align}
\Omega_{i j}(T) \equiv - U_i \widetilde{\Lambda}_{i j}(0 ; T). 
\label{Omega T}
\end{align} 
The eigenvalues of \eqref{Omega T} can then be called \emph{non-equilibrium magnon frequencies}, and they are time-dependent due to the action of the external field. It should be noted that the system given by the union of the magnetic medium and the external field might in general have a lower spatial symmetry than the lattice of the magnetic medium in the absence of the field (the field typically has some privileged directions, such as the polarization and direction of propagation for an electromagnetic wave). If such symmetry lowering is absent or negligible, one can exploit the symmetry of the magnetic lattice to diagonalize $\Omega_{i j}(T)$ [see Appendix \ref{app: periodic}]. 

In equilibrium, which is formally a particular case of this treatment which is obtained when the Hamiltonian is time-independent, $\Omega_{i j}$ is independent of $T$ and its eigenvalues are the conventional magnon frequencies. Therefore, we have formally demonstrated that the minimal correction to the transverse magnetic susceptibility in non-equilibrium situations, valid in the adiabatic regime, consists in the fact that the magnon frequencies acquire a time dependence. 

We note that the approximation which produces Eq.\eqref{Omega T}, namely replacing $\widetilde{\Lambda}_{i j}(\omega ; T) \rightarrow \widetilde{\Lambda}_{i j}(0 ; T)$, corresponds to linearizing the eigenvalue problem associated with Eq.\eqref{f matrix adiabatic}. Corrections can be computed by keeping into account higher-order terms in the Taylor expansion of $\widetilde{\Lambda}_{i j}(\omega ; T)$ in powers of $\omega$; such analysis is beyond the scope of this work.

We now characterize the non-equilibrium magnon frequencies and establish the correspondence to the previous literature, by introducing two different forms of non-equilibrium exchange parameters.

\section{Non-equilibrium exchange parameters}
\label{sec: exch}

\subsection{Two-times exchange parameters}

We first switch back from the frequency-domain representation to the time-domain representation. We define the two-times exchange matrix
\begin{align}
\Omega_{i j}(t, t') \stackrel{\mathrm{THF}}{=} \, \stackrel{\mathrm{SOH}}{=}   - U_i \Lambda_{i j}(t, t')  ,
\label{Omega tt'}
\end{align}
and we express it in terms of non-equilibrium 1EGFs and self-energies. To this end, we use the non-equilibrium Dyson equations in the THF approximation, Eqs.\eqref{Dyson THF SOH}, to eliminate the hopping matrix $T$ from the expression of $\Lambda$, Eq.\eqref{Lambda GF}. We obtain
\begin{align}
\Lambda_{i j}(t, t')  \stackrel{\mathrm{THF}}{=} \, & \mathrm{i} \theta(t - t') \left[ 2 \Sigma_{j S}(t') - \mathrm{i} \overrightarrow{\partial}_{t'} \right] \nonumber \\
& \times \left[ \left(   G^<_{\downarrow}  \right)^{i t}_{j t'}    \left(   G^>_{\uparrow}  \right)_{i t}^{j t'}   
-   \left(   G^>_{\downarrow}  \right)^{i t}_{j t'}    \left(   G^<_{\uparrow}  \right)_{i t}^{j t'}   \right] .
\end{align}
We split the exchange matrix into two parts,
\begin{align}
\Omega_{i j}(t, t') \stackrel{\mathrm{THF}}{=} \, \stackrel{\mathrm{SOH}}{=}   \frac{4}{m_i \left( \frac{t + t'}{2} \right)} \left[ J_{i j}(t, t') + X_{i j}(t, t') \right] ,
\end{align}
where
\begin{align}
  J_{i j}(t, t')  \equiv  \, &      \mathrm{i} \theta(t - t')  \, \Sigma_{i S}\!\left( \frac{t + t'}{2} \right)     \Sigma_{j S}(t')  \nonumber \\
  & \times    \left[    \left( G^<_{\downarrow} \right)_{j   t'}^{i   t} \!   \left( G^>_{\uparrow} \right)^{j   t'}_{i   t} 
-   \left( G^>_{\downarrow} \right)^{i   t}_{j   t'} \!   \left( G^<_{\uparrow} \right)_{i  t}^{j   t'}   
    \right]          
    \label{Jtt'}   
\end{align}
is the \emph{two-times exchange parameter} (equivalent to the analogous quantity obtained in Ref.\onlinecite{Secchi13}), and
\begin{align}
 X_{i j}(t, t') \equiv     &  \, \theta(t - t')   \,   \frac{1}{2} \Sigma_{i S} \! \left( \frac{t + t'}{2} \right)   \nonumber \\
 & \times \!  \overrightarrow{\partial}\!_{t'} \! \left[ \left( G^<_{\downarrow} \right)_{j  t'}^{i   t}  \!  \left( G^>_{\uparrow} \right)^{j   t'}_{i   t}  \! - \left( G^>_{\downarrow} \right)^{i   t}_{j   t'}    \left( G^<_{\uparrow} \right)_{i   t}^{j   t'}   
       \right]       
       \label{Xtt'}
\end{align}
is a quantity whose meaning will be clarified in Section \ref{one-time}. Switching again to the Wigner-coordinates representation and Laplace transforming with respect to relative time, we obtain
\begin{align}
\widetilde{\Omega}_{i j}(\omega ; T) \stackrel{\mathrm{THF}}{=} \, \stackrel{\mathrm{SOH}}{=}    \frac{4}{m_i\left( T  \right)}  \left[ \widetilde{J}_{i j}\left(  \omega ; T \right) + \widetilde{X}_{i j}(\omega; T)        \right]  . 
\label{time-dependent frequency matrix}
\end{align}
We simplify the second term in the RHS of Eq.\eqref{time-dependent frequency matrix}; after performing partial integration and using the relation $\left( G^> \right)^{i, T}_{j, T} = - \mathrm{i} \delta^i_j + \left( G^< \right)^{i, T}_{j, T}$, we obtain
\begin{align}
\widetilde{X}_{i j}(\omega; T)    =  & -   \frac{1}{2}  \delta_{ij} \Sigma_{i S}(T) \, m_i(T)    \nonumber \\
& +     \frac{1}{2}   \Sigma_{i S}(T) \left( \frac{1}{2} \overrightarrow{\partial}_T + \mathrm{i} \omega  \right)    \int_{0}^{\infty}  \! \mathrm{d} \tau \mathrm{e}^{\mathrm{i} \omega \tau}   \nonumber \\
& \times \Bigg[ \left( G^<_{\downarrow} \right)_{j , T - \tau/2 }^{i , T + \tau/2 }    \left( G^>_{\uparrow} \right)^{j  , T - \tau/2}_{i ,  T + \tau/2} \nonumber \\
&  -  \left( G^>_{\downarrow} \right)^{i ,   T + \tau/2 }_{j ,  T - \tau/2}   \left( G^<_{\uparrow} \right)_{i  , T + \tau/2}^{j  ,  T - \tau/2}       \Bigg]    .
\label{X adiabatic}
\end{align}
The first term in the RHS of Eq.\eqref{time-dependent frequency matrix} involves the Laplace transform of the two-times exchange parameters,
\begin{widetext}
\begin{align}
 \widetilde{J}_{i j}\left(  \omega ; T \right) \equiv \,  & \mathrm{i} \, \Sigma_{i S}( T )  \int_{0}^{\infty} \mathrm{d} \tau \mathrm{e}^{\mathrm{i} \omega \tau}        \,  \Sigma_{j S}( T - \tau/2)  
  \left[    \left( G^<_{\downarrow} \right)_{j ,  T - \tau/2}^{i ,  T + \tau/2}  
  \left( G^>_{\uparrow} \right)^{j ,  T - \tau/2}_{i ,  T + \tau/2} 
-   \left( G^>_{\downarrow} \right)^{i ,  T + \tau/2}_{j ,  T - \tau/2}   \left( G^<_{\uparrow} \right)_{i , T + \tau/2}^{j ,  T - \tau/2}    \right]  .             
\end{align}
\end{widetext}

\subsection{One-time exchange parameters}
\label{one-time}

If $\widetilde{\Omega}_{i j}(\omega ; T)$ is almost independent of $\omega$, we can determine a \emph{time-dependent} pole of the non-equilibrium transverse susceptibility, which is the generalization of the magnon frequency to the non-equilibrium adiabatic regime, as in Eq.\eqref{Omega T}. More explicitly, from Eq.\eqref{time-dependent frequency matrix} we write
\begin{align}
\Omega_{i j}(T) & \equiv    \lim_{\epsilon \rightarrow 0^+} \lim_{\omega \rightarrow 0} \widetilde{\Omega}_{i j}(\omega + \mathrm{i} \epsilon ; T) \nonumber \\
& \equiv \frac{4}{m_i\left( T  \right)}  \left[  J_{i j}\left(  T \right) +  X_{i j}(  T)        \right] , 
\label{time-dependent magnon frequencies}
\end{align} 
where $\omega$ and $\epsilon > 0$ are real, and  
\begin{widetext}
\begin{align}
J_{i j}(T)    = \, &  \mathrm{i} \, \Sigma_{i S}( T )  \lim_{\epsilon \rightarrow 0^+} \int_{0}^{\infty} \mathrm{d} \tau  \mathrm{e}^{ - \epsilon  \tau }        \,  \Sigma_{j S}( T -  \tau/2)   
 \left[    \left( G^<_{\downarrow} \right)_{j ,  T -  \tau/2}^{i ,  T +  \tau/2}  
  \left( G^>_{\uparrow} \right)^{j ,  T -  \tau/2}_{i ,  T +  \tau/2} 
-   \left( G^>_{\downarrow} \right)^{i ,  T +  \tau/2}_{j ,  T -  \tau/2}   \left( G^<_{\uparrow} \right)_{i , T +  \tau/2}^{j ,  T -  \tau/2}    \right] ,
\label{J EXCHANGE}
\end{align}
\begin{align}
 X_{i j}(T)    =  & -   \frac{1}{2}  \delta_{ij} \Sigma_{i S}(T) \, m_i(T)    \nonumber \\
& +     \frac{1}{4}   \Sigma_{i S}(T)  \lim_{\epsilon \rightarrow 0^+}  \overrightarrow{\partial}_T      \int_{0}^{\infty}  \! \mathrm{d} \tau \mathrm{e}^{- \epsilon \tau}    \left[ \left( G^<_{\downarrow} \right)_{j , T - \tau/2 }^{i , T + \tau/2 }    \left( G^>_{\uparrow} \right)^{j  , T - \tau/2}_{i ,  T + \tau/2}    -  \left( G^>_{\downarrow} \right)^{i ,   T + \tau/2 }_{j ,  T - \tau/2}   \left( G^<_{\uparrow} \right)_{i  , T + \tau/2}^{j  ,  T - \tau/2}       \right]    .
\label{X EXCHANGE}
\end{align}
\end{widetext}
As seen in Eq.\eqref{time-dependent magnon frequencies}, both terms $J_{i j}(T)$ and $X_{i j}(T)$ contribute on the same footing to the time-dependent magnon dispersion. We identify $J_{i j}(T)$ given in Eq.\eqref{J EXCHANGE} as the time-dependent exchange parameter due to its non-locality in space and its general structure that can be schematically denoted as $\Sigma G \Sigma G$, which is analogous to the structure found for the equilibrium exchange parameters in equilibrium theories (see e.g. Refs.\onlinecite{Katsnelson00, Katsnelson02, Secchi16}). The term $X_{i j}(T)$ defined in Eq.\eqref{X EXCHANGE} is given by two contributions. The first line is local in space; an analogous term appears in the expression of the dynamical transverse susceptibility in equilibrium (see Section \ref{section recover equilibrium}), of which this is the non-equilibrium generalization. The second line is a purely (non-local) non-equilibrium term with no analogue in equilibrium. In fact, the Green's functions would not depend on $T$ in that case, so the derivative would vanish. Out of equilibrium, instead, the $T$ dependence is not trivial, due to the time-dependent hopping. This term is explicitly related to the dynamical variation of the sites' electronic population. The presence of the term $X_{i j}(T)$ in the expression of the susceptibility has an important role in showing that the magnon dispersion satisfies the Goldstone theorem, even out of equilibrium (see Section \ref{sec: Goldstone}).

\subsection{Equilibrium exchange parameters} 
\label{section recover equilibrium}

The equilibrium regime is a particular case of the adiabatic regime, such that 1EGFs depend only on the relative time $\tau$ and not on the total time $T$, while THF self-energies are time-independent. The equilibrium exchange parameters are obtained from Eqs.\eqref{J EXCHANGE} and \eqref{X EXCHANGE} by removing the dependence on $T$. If the state of the system is given by a thermal distribution, in the limit of zero temperature (or inverse temperature $\beta \rightarrow \infty$) we can apply the analytical continuation from the real-time branches of the KP contour to the imaginary-time branch, and represent 1EGFs in the Matsubara formalism. In this case, we obtain [details are given in Appendix \ref{app: equilibrium}]
\begin{align}
&  J_{i j}    =   \frac{1}{2}    \lim_{\beta \rightarrow \infty} \frac{1}{\beta} \sum_{\omega_n} \Delta_i \,  G_{i j}^{\downarrow}(\mathrm{i} \omega_n) \,  \Delta_j  \, G_{j i}^{\uparrow}(\mathrm{i} \omega_n)     , \nonumber \\
& X_{i j} =    \frac{1}{4}  \delta_{ij} \Delta_{i}   m_i .
 \label{J Matsubara}
\end{align}
This result agrees with the equilibrium formulas derived with different methods in Refs. \onlinecite{Katsnelson00, Katsnelson02, Secchi13, Secchi15, Secchi16}, specialized to the SOH model in the HF approximation. We see that Eq.\eqref{X EXCHANGE} is the non-equilibrium generalization of the last term of Eq.(31) in Ref.\onlinecite{Katsnelson04}.

\section{Goldstone theorem} 
\label{sec: Goldstone}

The SOH model is not relativistic, therefore rotating all the electronic spins of the same angle with respect to a given axis costs no energy. Since this is a continuous symmetry, the Goldstone theorem predicts that the exchange matrix has a null eigenvalue, which in a lattice corresponds to the eigenstate with $\boldsymbol{q} = \boldsymbol{0}$ (that is, $\lim_{\boldsymbol{q} \rightarrow \boldsymbol{0}}\omega_{\boldsymbol{q}} = 0$). We recover this result in our theory, even out of equilibrium, since it immediately follows from Eqs.\eqref{Lambda GF} and \eqref{Omega tt'} that
\begin{align}
\sum_{j} \Lambda_{i j}(t, t') = 0 \Rightarrow \sum_j \Omega_{i j}(t, t') = 0 ,
\label{Goldstone satisfied}
\end{align}     
hence the vector $(1, 1, 1, \ldots, 1)$ is an eigenvector of the exchange matrix $\Omega(t, t')$, with eigenvalue $\omega = 0$ (if the system is a lattice, such eigenvector corresponds indeed to the state with $\boldsymbol{q} = \boldsymbol{0}$). Obviously, this property holds also in equilibrium, as a particular case. An alternative way to check that our theory is consistent with the Goldstone theorem is shown in Appendix \ref{app: sum rule}. 

The Goldstone theorem suggests a possible alternative definition for the exchange parameters contributing to the (one-time) exchange matrix. We can define \emph{starred} exchange parameters by combining Eq.\eqref{J EXCHANGE} and the non-local part of Eq.\eqref{X EXCHANGE} (second line). We get:
\begin{widetext}
\begin{align}
J^{\star}_{i j}(T)    = \, &  \mathrm{i} \, \Sigma_{i S}( T )  \lim_{\epsilon \rightarrow 0^+} \int_{0}^{\infty}\!  \mathrm{d} \tau  \mathrm{e}^{ - \epsilon  \tau }     \!   \left[  \Sigma_{j S}( T -  \tau/2)    -     \frac{\mathrm{i}}{4}        \overrightarrow{\partial}_T  \right] \!\!
 \left[    \left( G^<_{\downarrow} \right)_{j ,  T -  \tau/2}^{i ,  T +  \tau/2}   \!
  \left( G^>_{\uparrow} \right)^{j ,  T -  \tau/2}_{i ,  T +  \tau/2} 
-   \left( G^>_{\downarrow} \right)^{i ,  T +  \tau/2}_{j ,  T -  \tau/2} \!  \left( G^<_{\uparrow} \right)_{i , T +  \tau/2}^{j ,  T -  \tau/2}    \right]         .
\end{align}
\end{widetext}
Combining this definition with Eq.\eqref{Goldstone satisfied}, we can re-write Eq.\eqref{time-dependent magnon frequencies} in terms of $J^{\star}$ only as
\begin{align}
\Omega_{i j}(T)   \equiv \frac{4}{m_i\left( T  \right)}  \left[  J^{\star}_{i j}\left(  T \right) - \delta_{i j} \sum_k J^{\star}_{i k}(  T)        \right]  .
\end{align}

\section{Summary} 
\label{sec: summary}

To summarize, we have presented a rigorous derivation of the transverse spin susceptibility in the non-equilibrium adiabatic regime for the SOH model within the THF approximation, leading to the definition of non-equilibrium magnon frequencies and exchange parameters. Our results should be relevant to interpret the physics associated with ultrafast laser experiments, and possibly to unravel the effect of phonons on the magnetic properties of materials, provided that the frequencies of the oscillating fields are much smaller than the Stoner excitations. Further work can be envisaged to remove the THF approximation and extend to more general electronic systems, including relativistic interactions. The starting point for these possible developments is given by the GBSE, Eq.\eqref{BS t'}.

Concerning the possibility of developing a non-equilibrium theory beyond the THF approximation, we mention that using exact Green’s functions but neglecting the vertices is not acceptable because it would break the Goldstone theorem \cite{Hertz73}. The possibility of obtaining a problem that can be solved in closed form without employing the THF approximation must rely on the assumption of some small parameter (and therefore, a necessary loss of generality with respect to the unspecified electronic configuration that we have considered here). In equilibrium, a technique involving exact Green's functions of the Hubbard $X$-operators was presented in Refs.\onlinecite{Irkhin85a, Irkhin85b}, applied to study a fully spin-polarized electronic system with a small concentration of holes, with emphasis on the two-magnon scattering processes. The inclusion of full Green’s functions beyond Hartree-Fock was possible due to the assumed smallness of either the concentration of holes, or the inverse number of nearest neighbours, which allowed a linear approximation in one of those parameter. The generalization of this technique to the non-equilibrium regime is beyond the scope of the present work, where we have instead focused on obtaining the THF-approximated results without making any assumption on the electronic configuration.

\section*{Acknowledgements}

A. S. and M. I. K. acknowledge funding from the Nederlandse Organisatie voor Wetenschappelijk Onderzoek (NWO) via a Spinoza Prize; A. I. L. acknowledges funding from the SFB-925 grant of Deutsche Forschungsgemeinschaft (DFG). We thank Mikhail Titov and Alessandro Principi for useful discussions.

\appendix

\section{Derivation of the generalized Bethe-Salpeter equation}
\label{app: GBSE}

We derive the generalized Bethe-Salpeter equation using the properties of the non-equilibrium Green functions. The Dyson equations on the KP contour are written as 
\begin{align}
& \sum_3 \int_{\gamma} \mathrm{d} z_3 \left( \overrightarrow{G^{-1}}_{\boldsymbol{B}} \right)^{1 z_1}_{3 z_3} \left( G_{\boldsymbol{B}} \right)^{3 z_3}_{2 z_2} = \delta^1_2 \delta(z_1, z_2), \nonumber \\
& \sum_3 \int_{\gamma} \mathrm{d} z_3  \left( G_{\boldsymbol{B}} \right)^{1 z_1}_{3 z_3}  \left( \overleftarrow{G^{-1}}_{\boldsymbol{B}} \right)^{3 z_3}_{2 z_2}  = \delta^1_2 \delta(z_1, z_2)  ,
\end{align}
where
\begin{align}
& \left( \overrightarrow{G^{-1}}_{\boldsymbol{B}} \right)^{1 z_1}_{3 z_3} = \delta^1_3 \delta(z_1, z_3) \,  \mathrm{i} \overrightarrow{\partial}_{z_3}   - \left( T_{\boldsymbol{B}} \right)^{1 z_1}_{3 z_3}    - \left( \Sigma_{\boldsymbol{B}} \right)^{1 z_1}_{3 z_3}  , \nonumber \\
& \left( \overleftarrow{G^{-1}}_{\boldsymbol{B}} \right)^{3 z_3}_{2 z_2} = -    \mathrm{i} \overleftarrow{\partial}_{z_3} \delta^3_2 \delta(z_3, z_2)   - \left( T_{\boldsymbol{B}} \right)^{3 z_3}_{2 z_2}    - \left( \Sigma_{\boldsymbol{B}} \right)^{3 z_3}_{2 z_2} .
\end{align}
Here $\Sigma_{\boldsymbol{B}}$ and $T_{\boldsymbol{B}}$ denote, respectively, the self-energy and single-particle Hamiltonian matrix in the presence of a magnetic field $\boldsymbol{B}$ depending on the KP coordinate. In particular,
\begin{align}
\left(  T_{\boldsymbol{B}} \right)^{1 z_1}_{3 z_3} \equiv \delta(z_1 , z_3) \left\{ \left[T(z_1) \right]^{1}_{3}  +    \delta^{i_1}_{i_3}  \boldsymbol{B}_{i_1   z_1} \! \cdot \left( \boldsymbol{S}_{i_1} \right)^{M_1}_{M_3}  \right\},
\end{align}
where $\left[T(z_1) \right]^1_3$ is the hopping term that does not depend on $\boldsymbol{B}$, but is time-dependent as well, since it includes all the external fields acting on the electrons. Using a condensed notation, where the sums over all matrix indices and integrations over intermediate times are implied, we can write
\begin{align}
&       \overrightarrow{G^{-1}}_{\boldsymbol{B}} \cdot  G_{\boldsymbol{B}}   = 1 
\Rightarrow   \frac{\delta \overrightarrow{G^{-1}}_{\boldsymbol{B}}   }{\delta B_{j  z_3}^{\alpha'}} \cdot  G_{\boldsymbol{B}}    +    \overrightarrow{G^{-1}}_{\boldsymbol{B}} \cdot \frac{\delta G_{\boldsymbol{B}}   }{\delta B_{j  z_3}^{\alpha'}}    = 0  \nonumber \\
& \Rightarrow       G_{\boldsymbol{B}} \cdot \overrightarrow{G^{-1}}_{\boldsymbol{B}} \cdot \frac{\delta G_{\boldsymbol{B}}   }{\delta B_{j z_3}^{\alpha'}}    = -   G_{\boldsymbol{B}} \cdot \frac{\delta \overrightarrow{G^{-1}}_{\boldsymbol{B}}   }{\delta B_{j z_3}^{\alpha'}} \cdot  G_{\boldsymbol{B}}    .
\end{align}
We can replace $G_{\boldsymbol{B}} \cdot \overrightarrow{G^{-1}}_{\boldsymbol{B}} \rightarrow G_{\boldsymbol{B}} \cdot \overleftarrow{G^{-1}}_{\boldsymbol{B}} \equiv 1$, since the two expressions differ only by boundary terms which vanish due to the Kubo-Martin-Schwinger relations \cite{KMS relations} on the KP contour. We then obtain the identity
\begin{align}
      \frac{\delta G_{\boldsymbol{B}}   }{\delta B_{j  z_3}^{\alpha'}}      =    G_{\boldsymbol{B}} \cdot \frac{\delta \Sigma_{\boldsymbol{B}}   }{\delta B_{j  z_3}^{\alpha'} } \cdot  G_{\boldsymbol{B}}  + G_{\boldsymbol{B}} \cdot \frac{\delta T_{\boldsymbol{B}}   }{\delta B_{j  z_3}^{\alpha'} } \cdot  G_{\boldsymbol{B}}.
      \label{for BetheSalpeter}
\end{align}
We apply Eq.\eqref{for BetheSalpeter} to Eq.\eqref{most general chi}, obtaining  
\begin{align}
\chi^{\alpha \alpha'}_{1, 2 ; j }(z_1, z_2 ; z_3) \equiv & \left( \chi_0 \right)^{\alpha \alpha'}_{1, 2 ; j }(z_1, z_2 ; z_3) \nonumber \\
&  + \left( \chi_{\Gamma} \right)^{\alpha \alpha'}_{1, 2 ; j }(z_1, z_2 ; z_3) ,
\label{structure Bethe}
\end{align}
where
\begin{align}
& \left( \chi_0 \right)^{\alpha \alpha'}_{1, 2 ; j }(z_1, z_2 ; z_3)   \equiv - \mathrm{i} \left( S^{\alpha}   \cdot G_{j z_3}  \cdot S^{\alpha'}_{j}   \cdot  G^{j z_3} \right)^{1 z_1}_{2 z_2} \nonumber \\
&   = - \mathrm{i}  \sum_M  \left( S^{\alpha}   \cdot G   \cdot S^{\alpha'}  \right)^{1 z_1}_{j M z_3}    G^{j M z_3}_{2 z_2} ,
\label{chi_0 general}
\end{align}
\begin{align}
& \left( \chi_{\Gamma} \right)^{\alpha \alpha'}_{1, 2 ; j }(z_1, z_2 ; z_3)  \equiv  -\mathrm{i} \left(  S^{\alpha}  \cdot G   \cdot \left. \frac{\delta \Sigma_{\boldsymbol{B}}   }{\delta B_{j z_3}^{\alpha'}}   \right|_{\boldsymbol{B} = \boldsymbol{0}}   \cdot   G \right)^{1 z_1}_{2 z_2}   \nonumber \\
& =  -\mathrm{i} \int_{\gamma} \mathrm{d} (w_4 , w_5)  \sum_{4, 5} \left(  S^{\alpha}  \cdot G_{4 w_4}  \right)^{1 z_1} G^{5 w_5}_{2 z_2}  \nonumber \\
& \quad \times   \left. \frac{\delta \left( \Sigma_{\boldsymbol{B}} \right)^{4 w_4}_{5 w_5}   }{\delta B_{j z_3}^{\alpha'}}   \right|_{\boldsymbol{B} = \boldsymbol{0}}    .
\label{chi_Gamma general}
\end{align}

We now perform some manipulations on Eq.\eqref{chi_Gamma general}. If the dimensionality of the spin associated with quantum numbers $k$ is $\mathcal{S}_{k}$, then a fundamental property of the spin matrices is that 
\begin{align}
\left( \sum_{\alpha''} S_{k}^{\alpha''} \cdot S_{k}^{\alpha''} \right)^M_{M'} = \delta^M_{M'} \mathcal{S}_{k} \left( \mathcal{S}_{k} + 1 \right)  ,
\label{spin matrices sum rule} 
\end{align}
with $\alpha'' \in \lbrace x, y, z \rbrace$. Using this relation, we obtain
\begin{align}
&   \frac{\delta \left( \Sigma_{\boldsymbol{B}} \right)^{4 w_4}_{5 w_5}   }{\delta B_{j z_3}^{\alpha'}}    =  \int_{\gamma} \mathrm{d} (w_6, w_7)  \sum_{6 , 7}  \frac{\delta   \Sigma^{4 w_4}_{5 w_5}   }{\delta G^{6 w_6}_{7 w_7} }       \frac{\delta \left( G_{\boldsymbol{B}} \right)^{6 w_6}_{7 w_7}   }{\delta B_{j z_3}^{\alpha'}}     \nonumber \\
& = \int_{\gamma} \mathrm{d} (w_6 , w_7 )  \sum_{6 , 7} \frac{1}{\mathcal{S}_{i_4} \left( \mathcal{S}_{i_4} + 1 \right) \mathcal{S}_{i_6} \left( \mathcal{S}_{i_6} + 1 \right) }  \nonumber \\
& \quad  \times   \sum_{\alpha''} \frac{\delta \left( S^{\alpha''}_{i_4} \cdot S^{\alpha''}_{i_4} \cdot \Sigma \right)^{4 w_4}_{5 w_5}   }{\delta G^{6 w_6}_{7 w_7} }     \sum_{\alpha'''}     \frac{\delta \left( S_{i_6}^{\alpha'''}  \cdot S_{i_6}^{\alpha'''} \cdot G_{\boldsymbol{B}} \right)^{6 w_6}_{7 w_7}   }{\delta B_{j z_3}^{\alpha'}}   \nonumber \\
& = \int_{\gamma} \mathrm{d} (w_6 , w_7 )  \sum_{6 , 7} \frac{1}{\mathcal{S}_{i_4} \left( \mathcal{S}_{i_4} + 1 \right) \mathcal{S}_{i_6} \left( \mathcal{S}_{i_6} + 1 \right) }   \nonumber \\
& \quad \times  \sum_{\alpha''} \sum_{M'}  \left( S_{i_4}^{\alpha''} \right)_{M_4}^{M'} \frac{\delta \left(   S^{\alpha''}  \cdot \Sigma \right)^{4 w_4}_{5 w_5}   }{\delta G^{6 w_6}_{7 w_7} }  \nonumber \\
& \quad \times  \sum_{\alpha'''} \sum_M    \left( S_{i_6}^{\alpha'''} \right)_{M_6}^M    \frac{\delta \left(    S^{\alpha'''} \cdot G_{\boldsymbol{B}} \right)^{6  w_6}_{7 w_7}   }{\delta B_{j z_3}^{\alpha'}}   .  
\end{align}
Inserting this into Eq.\eqref{chi_Gamma general} yields
\begin{align}
& \left( \chi_{\Gamma} \right)^{\alpha \alpha'}_{1, 2 ; j }(z_1, z_2 ; z_3) \nonumber \\
&   \equiv  \sum_{\alpha''  \alpha'''}  \int_{\gamma} \mathrm{d} ( w_4 , w_5 , w_6 , w_7 )  \sum_{4 5 6 7}  \left( \chi_0^{\alpha \alpha''} \right)^{1 z_1, 5 w_5}_{2 z_2  , 4 w_4 }   \nonumber \\
& \quad \times  \left( \Gamma^{\alpha'' \alpha'''} \right)^{4 w_4, 7w_7}_{5 w_5, 6 w_6} \chi^{\alpha''' \alpha'}_{6, 7 ; j}(w_6, w_7 ; z_3)  ,
\label{chi_Gamma elaboration 1}
\end{align}
where we have introduced the quantities defined in Eqs.\eqref{the chi_0} and \eqref{the Gamma} of the main text. We note here that a more explicit form of Eq.\eqref{the Gamma} is
\begin{align}
 \left( \Gamma^{\alpha'' \alpha'''} \right)^{4 w_4, 7w_7}_{5 w_5, 6 w_6}      = &   \frac{\mathrm{i} }{\mathcal{S}_{i_4} \left( \mathcal{S}_{i_4} + 1 \right) \mathcal{S}_{i_6} \left( \mathcal{S}_{i_6} + 1 \right) }   \nonumber \\
& \times \! \! \sum_{ M  M'}  \!  \left( S_{i_4}^{\alpha''}  \right)^{M_4}_{M'} \frac{\delta   \Sigma^{i_4 M' w_4}_{5 w_5}   }{\delta G^{i_6 M w_6}_{7 w_7} }   \! \left( S_{i_6}^{\alpha'''} \right)_{M_6}^M  . 
\end{align}
Equation \eqref{chi_0 general} is related to Eq.\eqref{the chi_0} via the identity 
\begin{align}
& \left( \chi_0 \right)^{\alpha \alpha'}_{1, 2 ; j }(z_1, z_2 ; z_3)     =    \sum_M  \left( \chi_0^{\alpha \alpha'} \right)^{1 z_1,  \,  j M z_3}_{ 2 z_2 , \, j M z_3   } ,
\end{align}
and the quantity defined in Eq.\eqref{main chi_0} is related to Eq.\eqref{chi_0 general} via
\begin{align}
& \left( \chi_0 \right)^{\alpha \alpha'}_{1, 2 ; j }(z_1, z_2 ; t') \nonumber \\
& \equiv \left( \chi_0 \right)^{\alpha \alpha'}_{1, 2 ; j }(z_1, z_2 ; t'_{(+)})   - \left( \chi_0 \right)^{\alpha \alpha'}_{1, 2 ; j }(z_1, z_2 ; t'_{(-)})  .
\end{align}
By inserting Eq.\eqref{chi_Gamma elaboration 1} into Eq.\eqref{structure Bethe}, one obtains the generalized Bethe-Salpeter equation, given by Eq.\eqref{BS t'} of the main text.

\section{Non-equilibrium dynamical spin susceptibility}
\label{app: dynsusc}

In order to establish the relation between the super-matrix defined in Eq.\eqref{most general chi} and the physical susceptibility defined in Eq.\eqref{def chi} of the main text, it is first convenient to define the quantity
\begin{align}
\chi^{\alpha \alpha'}_{i j}(z_1, z_2 ; z_3) \equiv \sum_M  \chi^{\alpha \alpha'}_{i M, i M ; j }(z_1, z_2 ; z_3) .
\label{generalized chi}
\end{align}
We obtain the physical susceptibility from Eqs.\eqref{most general chi} and \eqref{generalized chi} as follows. From Eq.\eqref{generalized chi} we get
\begin{align}
& \sum_j \sum_{\alpha'} \int_{\gamma} \mathrm{d} z_3  \chi^{\alpha \alpha'}_{i j}(z_1, z_2 ; z_3) \delta B^{\alpha'}_{j z_3}  \nonumber \\
& = - \mathrm{i} \,  \sum_j \sum_{\alpha'} \int_{\gamma} \mathrm{d} z_3 \, \mathrm{Sp}  \left\{  S^{\alpha}_i \cdot  \left. \frac{\delta    \left( G_{\boldsymbol{B}}\right)^{i z_1}_{i z_2} }{\delta B_{j  z_3}^{\alpha'}} \right|_{\boldsymbol{B} = \boldsymbol{0}} \right\} \delta B^{\alpha'}_{j  z_3}  \nonumber \\
&   = \delta \left[ - \mathrm{i} \,  \mathrm{Sp} \left(  S^{\alpha}_i \cdot    G^{i z_1}_{i z_2} \right) \right] .
\end{align}
This quantity is equal to the variation of the local magnetic moment under a variation of the magnetic field, $\delta \left< \hat{S}^{\alpha}_i(t) \right>$, if we take $z_1 = t_{(+)} $ and $z_2 = t_{(-)}$. Moving from the KP coordinates to physical times ($z_3 \rightarrow t'_{(+)}$ if $z_3 \in \gamma_{+}$ and $z_3 \rightarrow t'_{(-)}$ if $z_3 \in \gamma_{-}$) gives
\begin{align}
& \delta \left< \hat{S}^{\alpha}_i(t) \right> \nonumber \\
& =   \sum_j \sum_{\alpha'} \int_{t_0}^{\infty} \mathrm{d} t' \Big[ \chi^{\alpha \alpha'}_{i j}(t_{(+)}, t_{(-)} ; t'_{(+)}) \, \delta B^{\alpha'}_{j  t'_{(+)}  }  \nonumber \\
&  \quad - \chi^{\alpha \alpha'}_{i j}(t_{(+)}, t_{(-)} ; t'_{(-)}) \, \delta B^{\alpha'}_{j  t'_{(-)}  }   \Big]  ,
\end{align}
where we have put $\delta B^{\alpha'}_{j  z  } = 0$ if $z \in \gamma_{\mathrm{M}}$. Moreover, the variation of the magnetic field is physically meaningful only if $\delta B^{\alpha'}_{j  t'_{(+)}} = \delta B^{\alpha'}_{j  t'_{(-)}} \equiv \delta B^{\alpha'}_j(t')$. This gives
\begin{align}
   \delta \left< \hat{S}^{\alpha}_i(t) \right>   =   \sum_j \sum_{\alpha'} \int_{t_0}^{\infty} \mathrm{d} t' \chi^{\alpha \alpha'}_{i j}(t , t')  \, \delta B^{\alpha'}_j(t' ) ,
\end{align}
where the physical susceptibility is obtained as
\begin{align}
\chi^{\alpha \alpha'}_{i j}(t , t')   \equiv    \chi^{\alpha \alpha'}_{i j}(t_{(+)}, t_{(-)} ; t'_{(+)})    - \chi^{\alpha \alpha'}_{i j}(t_{(+)}, t_{(-)} ; t'_{(-)}) .
\end{align}
Using Eq.\eqref{generalized chi}, we immediately obtain that the relation between Eq.\eqref{def chi} and Eq.\eqref{most general chi} is given by Eq.\eqref{physical} of the main text.

\section{Simplification of the Bethe-Salpeter equation in the case of the single-orbital Hubbard model}
\label{app: BSE Hub}

We here show the details of the simplification of the THF Bethe-Salpeter equation for transverse susceptibility in the single-orbital Hubbard model (SOH). Using the fact that $\left[ \hat{H}(t), \hat{S}^z \right] = 0$, so the total third component of the spin of the system is a good quantum number, we obtain
\begin{align}
& \chi^{\alpha''' -}_{i M'', i M' ; j }(t ; t')   \nonumber \\
& \stackrel{\mathrm{SOH}}{=}  - \mathrm{i} \theta(t - t') \sum_M \left( S^{\alpha'''}  \right)^{M''}_M \delta^M_{\downarrow} \delta^{\uparrow}_{M'} \left< \left[  \hat{\rho}^{i \downarrow}_{i \uparrow}(t) , \hat{S}^-_j(t')  \right]  \right> \nonumber \\
& =   - \mathrm{i} \theta(t - t') \delta^{\uparrow}_{M'}  \left( S^{\alpha'''}  \right)^{M''}_{\downarrow}    \left< \left[  \hat{S}^{+}_{i  }(t) , \hat{S}^-_j(t')  \right]  \right>  \nonumber \\
&   =     \delta^{\uparrow}_{M'}  \left( S^{\alpha'''}  \right)^{M''}_{\downarrow}   \chi^{+ -}_{i  j}(t, t') .
\end{align}
Equation \eqref{bare susceptibility matrix definition} simplifies as
\begin{align}
  \left( \chi_0 \right)^{+ -}_{1, 2 ; j }(t , t')     \stackrel{\mathrm{SOH}}{=} & - \mathrm{i}  \theta(t - t') \delta^{M_1}_{\uparrow}      \Big[   \left( G^> \right)^{i_1 \downarrow t}_{j \downarrow t'}      \left(G^< \right)^{j \uparrow t' }_{2 t }  
 \nonumber \\
&    - \left( G^< \right)^{i_1 \downarrow t}_{j \downarrow t'}   \left( G^> \right)^{j \uparrow t' }_{2 t }   \Big]  ,
\end{align}
from which
\begin{align}
  \left( \chi_0 \right)^{+ - }_{i   j }(t , t')     \stackrel{\mathrm{SOH}}{=} &   - \mathrm{i}  \theta(t - t')        \Big[   \left( G^> \right)^{i \downarrow t}_{j \downarrow t'}      \left(G^< \right)^{j \uparrow t' }_{i \uparrow t }   \nonumber \\
&   - \left( G^< \right)^{i  \downarrow t}_{j \downarrow t'}   \left( G^> \right)^{j \uparrow t' }_{i \uparrow t }   \Big]  .
\end{align}
Equation \eqref{another one} simplifies as
\begin{align}
&  \chi_0^{+ \alpha''}\!(t, t'')^{ i M , 5   }_{ i M   , 4   } \nonumber \\
&  \stackrel{\mathrm{SOH}}{=} - \mathrm{i} \theta(t - t'') \delta^M_{\uparrow} \delta^{M_5}_{\uparrow} \left( S^{\alpha''}  \right)^{\downarrow}_{M_4} \Big[    \left( G^{>} \right)^{i \downarrow t}_{i_4 \downarrow t''}    \left( G^< \right)^{i_5 \uparrow t'' }_{i \uparrow  t }  \nonumber \\
& \quad \quad   - \left( G^{<} \right)^{i \downarrow t}_{i_4 \downarrow t''}    \left( G^> \right)^{i_5 \uparrow t'' }_{i \uparrow  t } \Big]    \nonumber \\
& \equiv   \delta^M_{\uparrow} \delta^{M_5}_{\uparrow} \left( S^{\alpha''}  \right)^{\downarrow}_{M_4}  \left(\chi_0 \right)^{+ -}_{i, i_4 i_5}(t, t'') .
\end{align}
Using these expressions, from Eq.\eqref{BSalp HF 2} one obtains Eq.\eqref{BSalp THF SOH} of the main text.

\section{Derivation of the equations for the susceptibility in the frequency domain}
\label{app: frequency derivation}

We here show the detailed derivation of the frequency-domain representations of the Bethe-Salpeter equation, Eq.\eqref{BSalp THF SOH}, and the equation for the bare susceptibility, Eq.\eqref{bare susceptibility i j t t'}.

We start from the Bethe-Salpeter equation. Equation \eqref{BSalp THF SOH} becomes
\begin{align}
&  \widetilde{\chi}^{+ -}_{i   j }(\tau , T ) - \left( \widetilde{\chi}_0 \right)^{+-}_{i   j }(\tau , T) \nonumber \\
 & \stackrel{\mathrm{THF}}{=} \, \stackrel{\mathrm{SOH}}{=}         \sum_{k}   \int_{- \infty}^{\infty} \mathrm{d}t''  \left( \widetilde{\chi}_0 \right)^{+ -}_{i  k}\! \left( T + \frac{\tau}{2} - t'', \frac{ T }{2} + \frac{\tau}{4} + \frac{t''}{2} \right)   \nonumber \\
  & \quad \quad \quad \quad \times  \left( - U_k \right)  \widetilde{\chi}^{+ -}_{k   j } \left(t'' - T + \frac{\tau}{2} , \frac{t''}{2} + \frac{T}{2} - \frac{\tau}{4} \right) ,
\end{align}
where we have extended the lower boundary of integration over $t''$ to $-\infty$. Applying the Laplace and Fourier transforms, we get
\begin{align}
&   \widetilde{\chi}^{+ -}_{i   j }(\omega , \Omega ) - \left( \widetilde{\chi}_0 \right)^{+-}_{i   j }(\omega , \Omega)   \nonumber \\
& \stackrel{\mathrm{THF}}{=} \, \stackrel{\mathrm{SOH}}{=}         \int_0^{\infty} \mathrm{d} \tau \, \mathrm{e}^{\mathrm{i} \omega \tau} \int_{-\infty}^{\infty}   \mathrm{d} T  \, \mathrm{e}^{\mathrm{i} \Omega T}  \sum_{k}   \int_{- \infty}^{\infty} \mathrm{d}t''   \nonumber \\
& \quad \quad \quad \quad \times \left( \widetilde{\chi}_0 \right)^{+ -}_{i  k}\left( T + \frac{\tau}{2} - t'', \frac{ T }{2} + \frac{\tau}{4} + \frac{t''}{2} \right)     \nonumber \\
 &  \quad \quad \quad \quad  \times    \left( - U_k \right) \widetilde{\chi}^{+ -}_{k   j }\left(t'' - T + \frac{\tau}{2} , \frac{t''}{2} + \frac{T}{2} - \frac{\tau}{4} \right) .
\end{align}
We change variables according to $t'' = \tau' - \frac{\tau}{2} + T$,
\begin{align}
&   \widetilde{\chi}^{+ -}_{i   j }(\omega , \Omega ) - \left( \widetilde{\chi}_0 \right)^{+-}_{i   j }(\omega , \Omega)  \nonumber \\
   & \stackrel{\mathrm{THF}}{=} \, \stackrel{\mathrm{SOH}}{=}      -  \int_0^{\infty} \mathrm{d} \tau \, \mathrm{e}^{\mathrm{i} \omega \tau} \int_{-\infty}^{\infty}   \mathrm{d} T  \, \mathrm{e}^{\mathrm{i} \Omega T}    \int_{- \infty}^{\infty} \mathrm{d}\tau' \nonumber \\
&   \quad \quad \quad \quad  \times \sum_{k} \left( \widetilde{\chi}_0 \right)^{+ -}_{i  k}\left(  \tau - \tau' , T + \frac{\tau'}{2} \right)        \nonumber \\
  &  \quad \quad \quad \quad   \times  U_k \, \widetilde{\chi}^{+ -}_{k   j }\left( \tau' , T - \frac{\tau - \tau'}{2}   \right) .
\end{align}
Using the inverse Fourier transform on the second arguments of the two susceptibilities, we perform the integration over $T$,
\begin{align}
&  \widetilde{\chi}^{+ -}_{i   j }(\omega , \Omega ) - \left( \widetilde{\chi}_0 \right)^{+-}_{i   j }(\omega , \Omega) \nonumber \\
& \stackrel{\mathrm{THF}}{=} \, \stackrel{\mathrm{SOH}}{=}          -  \int_0^{\infty} \mathrm{d} \tau      \int_{- \infty}^{\infty} \mathrm{d}\tau'  
  \int_{-\infty}^{\infty} \frac{\mathrm{d} \Omega'}{2 \pi}   \mathrm{e}^{\mathrm{i} \left( \omega + \frac{\Omega - \Omega'}{2} \right)  \tau}    \mathrm{e}^{  - \mathrm{i}      \frac{  \Omega }{2}  \tau' } 
   \nonumber \\
   & \quad \quad \quad \quad \times   \sum_{k}  \left( \widetilde{\chi}_0 \right)^{+ -}_{i  k}\left(  \tau - \tau' ;  \Omega' \right)       \, U_k \, \widetilde{\chi}^{+ -}_{k   j }\left( \tau' ; \Omega - \Omega'  \right)  .
\end{align}
For a fixed $\tau'$, we substitute $\sigma = \tau - \tau'$ and we obtain:
\begin{align}
 &  \widetilde{\chi}^{+ -}_{i   j }(\omega , \Omega ) - \left( \widetilde{\chi}_0 \right)^{+-}_{i   j }(\omega , \Omega)  \nonumber \\
& \stackrel{\mathrm{THF}}{=} \, \stackrel{\mathrm{SOH}}{=}          -   \int_{- \infty}^{\infty} \mathrm{d}\tau' \int_{- \tau'}^{\infty} \mathrm{d} \sigma       
  \int_{-\infty}^{\infty} \frac{\mathrm{d} \Omega'}{2 \pi}   \mathrm{e}^{\mathrm{i} \left( \omega + \frac{\Omega - \Omega'}{2} \right)    \sigma   } \mathrm{e}^{\mathrm{i} \left( \omega    - \frac{\Omega'}{2} \right)    \tau'   } \nonumber \\
  & \quad \quad \quad \quad  \times   \sum_{k}  \left( \widetilde{\chi}_0 \right)^{+ -}_{i  k}\left(  \sigma ;  \Omega' \right)           U_k \, \widetilde{\chi}^{+ -}_{k  j }\left( \tau' ; \Omega - \Omega'  \right)  .
\end{align}
Finally, we notice that the integrand vanishes when $\tau' < 0$ because $\widetilde{\chi}\left( \tau' ; \ldots  \right) \propto \theta(\tau')$, so we can restrict the integration over $\tau'$ to the interval $(0, \infty)$. The integrand also vanishes when $\sigma < 0$ because $\widetilde{\chi}_0\left( \sigma ; \ldots  \right) \propto \theta(\sigma)$, so we can also restrict the integration over $\sigma$ to the interval $(0, \infty)$. We then recognize two Laplace transforms, and we obtain
\begin{align}
  & \widetilde{\chi}^{+ -}_{i   j }(\omega , \Omega ) - \left( \widetilde{\chi}_0 \right)^{+-}_{i   j }(\omega , \Omega)   \nonumber \\
& \stackrel{\mathrm{THF}}{=} \, \stackrel{\mathrm{SOH}}{=}          -    
  \int_{-\infty}^{\infty} \frac{\mathrm{d} \Omega'}{2 \pi}        \sum_{k}  \left( \widetilde{\chi}_0 \right)^{+ -}_{i  k}\left(  \omega + \frac{\Omega - \Omega'}{2} ,  \Omega' \right)   \nonumber \\
& \quad \quad \quad \quad \times  U_k \, \widetilde{\chi}^{+ -}_{k   j }\left( \omega    - \frac{\Omega'}{2} , \Omega - \Omega'  \right)  .
\end{align}

We now treat the equation for the bare susceptibility, Eq.\eqref{bare susceptibility i j t t'}. Introducing the Wigner coordinates, we obtain
\begin{align}
& \left( \widetilde{\chi}_0 \right)^{+ -}_{i j}(\tau , T) \left[   \mathrm{i} \overleftarrow{\partial_{\tau}} - \frac{\mathrm{i}}{2} \overleftarrow{\partial_{T}}  - \Delta_j\left( T - \frac{\tau}{2} \right) \right]   \nonumber \\
& \stackrel{\mathrm{SOH}}{=} \, \stackrel{\mathrm{THF}}{=}  \delta( \tau ) \delta_{i j}  m_j(T) +  \widetilde{\Lambda}_{i j}( \tau , T)      .
\end{align}
We multiply both sides of the previous equation by $\mathrm{e}^{\mathrm{i} \omega \tau} \mathrm{e}^{\mathrm{i} \Omega T}$ and we integrate over $\tau$ and $T$, in both cases on the full real axis $(-\infty, \infty)$. We obtain:
\begin{align}
 &   \delta_{i j} m_j(\Omega) +  \widetilde{\Lambda}_{i j}( \omega , \Omega)  \nonumber \\
&    \stackrel{\mathrm{SOH}}{=} \, \stackrel{\mathrm{THF}}{=}   
\int_{- \infty}^{\infty} \mathrm{d} \tau \int_{- \infty}^{\infty} \mathrm{d} T \int_{-\infty}^{\infty} \frac{\mathrm{d} \Omega'}{2 \pi}  \mathrm{e}^{- \mathrm{i} \Omega' T} \nonumber \\
& \quad \quad \quad \quad \times \left( \widetilde{\chi}_0 \right)^{+ -}_{i  j}(\tau ; \Omega')   \left(   \mathrm{i} \overleftarrow{\partial_{\tau}} - \frac{\Omega'}{2}     \right) \mathrm{e}^{\mathrm{i} \omega \tau} \mathrm{e}^{\mathrm{i} \Omega T}  \nonumber \\
&   -      \int_{- \infty}^{\infty} \frac{\mathrm{d} \Omega'}{2 \pi}        \left( \widetilde{\chi}_0 \right)^{+ -}_{i  j}\!\left( \omega    +   \frac{\Omega - \Omega'}{2} ,  \Omega'  \right) \,  \Delta_j\!\left( \Omega - \Omega' \right)   .
\end{align}
We partially integrate on the variable $\tau$, and we note that the boundary terms vanish, respectively, because $\Im(\omega) > 0$ and $\left(\widetilde{\chi}_0 \right)^{+ -}_{i  j}(\tau ; \Omega') \propto \theta(\tau)$. We then obtain Eq.\eqref{bare susceptibility i j omega Omega} of the main text.

\section{Simplifications for spatially-periodic systems}
\label{app: periodic}

If the system is spatially periodic (and stays so under the application of the time-dependent external field), it is convenient to write and solve the equations for the susceptibility in wave-vector space. We define the spatial Fourier transforms according to the usual conventions,
\begin{align}
& f_{i  j} = \frac{1}{N} \! \sum_{\boldsymbol{q}} \mathrm{e}^{\mathrm{i} \boldsymbol{q} \cdot \left( \boldsymbol{R}_i - \boldsymbol{R}_j \right)} f_{\boldsymbol{q}} \Leftrightarrow  f_{\boldsymbol{q}} = \frac{1}{N} \sum_{i , j} \mathrm{e}^{ - \mathrm{i} \boldsymbol{q} \cdot \left( \boldsymbol{R}_i - \boldsymbol{R}_j \right)} f_{i  j } , \nonumber \\
& g_{i} = \frac{1}{\sqrt{N}} \sum_{\boldsymbol{q}} \mathrm{e}^{\mathrm{i} \boldsymbol{q} \cdot \boldsymbol{R}_i } g_{\boldsymbol{q}} \Leftrightarrow  g_{\boldsymbol{q}} = \frac{1}{\sqrt{N}} \sum_{i } \mathrm{e}^{ - \mathrm{i} \boldsymbol{q} \cdot \boldsymbol{R}_i } g_{i  } .
\end{align}
By applying $\frac{1}{N} \sum_{i , j} \mathrm{e}^{ - \mathrm{i} \boldsymbol{q} \cdot \left( \boldsymbol{R}_i - \boldsymbol{R}_j \right)}$ to both Eqs.\eqref{Bethe-Salpeter i j omega Omega} and \eqref{bare susceptibility i j omega Omega}, we obtain
\begin{widetext}
\begin{align}
  & \widetilde{\chi}^{+ -}_{\boldsymbol{q} }(\omega , \Omega ) - \left( \widetilde{\chi}_0 \right)^{+-}_{ \boldsymbol{q} }(\omega , \Omega)   \stackrel{\mathrm{THF}}{=} \, \stackrel{\mathrm{SOH}}{=}          -    \overline{U}
  \int_{-\infty}^{\infty} \frac{\mathrm{d} \Omega'}{2 \pi}           \left( \widetilde{\chi}_0 \right)^{+ -}_{ \boldsymbol{q} }\!\left(  \omega + \frac{\Omega - \Omega'}{2} ,  \Omega' \right)       \widetilde{\chi}^{+ -}_{ \boldsymbol{q} }\!\left( \omega    - \frac{\Omega'}{2} , \Omega - \Omega'  \right)  ,
  \label{Bethe-Salpeter q omega Omega}
\end{align}
\begin{align}
&  \left(   \omega - \frac{\Omega}{2}     \right)     \left( \widetilde{\chi}_0 \right)^{+ -}_{ \boldsymbol{q} }(\omega , \Omega)     -      \int_{- \infty}^{\infty} \frac{\mathrm{d} \Omega'}{2 \pi}        \left( \widetilde{\chi}_0 \right)^{+ -}_{\boldsymbol{q}}\!\left( \omega    +   \frac{\Omega - \Omega'}{2} ,  \Omega'  \right) \,  \overline{\Delta}\left( \Omega - \Omega' \right)      \stackrel{\mathrm{THF}}{=}  \, \stackrel{\mathrm{SOH}}{=}      2 \pi \delta(\Omega)  \overline{m} +  \widetilde{\Lambda}_{ \boldsymbol{q} }( \omega , \Omega)   ,
\label{bare susceptibility q omega Omega}      
\end{align}
\end{widetext}
where we have introduced the spatial averages
\begin{align}
\overline{g} \equiv \frac{1}{N} \sum_{i} g_i = \frac{g_{\boldsymbol{q} = \boldsymbol{0}}}{\sqrt{N}}   ,
\end{align}
and we have noticed that the average magnetic moment
\begin{align}
  \frac{1}{N} \sum_{i} m_i (T) \equiv   \overline{m}
\end{align}
is independent of time. If $U_i \rightarrow U$ is spatially uniform, as we shall assume, then also $U \overline{m} = \overline{\Delta}$ is time-independent, then $\overline{\Delta}\left( \Omega - \Omega' \right) \rightarrow 2 \pi \delta(\Omega - \Omega')  \overline{\Delta}$, and Eq.\eqref{bare susceptibility q omega Omega} can be solved without further approximations:
\begin{align}
\left( \widetilde{\chi}_0 \right)^{+ -}_{ \boldsymbol{q} }(\omega , \Omega)   \stackrel{\mathrm{THF}}{=}  \,    \stackrel{\mathrm{SOH}}{=}  \frac{ 2 \pi \delta(\Omega)  \overline{m}   +  \widetilde{\Lambda}_{ \boldsymbol{q} }( \omega , \Omega ) }{ \omega  - \frac{\Omega}{2}   - \overline{\Delta}  }  .
\label{bare susceptibility q omega T adiabatic}      
\end{align}
By inserting this result into Eq.\eqref{Bethe-Salpeter q omega Omega}, applying the adiabatic approximation and switching to the representation in terms of $\left( \omega ; T \right)$ we obtain
\begin{align}
  &  \widetilde{\chi}^{+ -}_{\boldsymbol{q} }(\omega ; T   )   \stackrel{\mathrm{THF}}{=} \, \stackrel{\mathrm{SOH}}{=}  \, \stackrel{\mathrm{AD}}{=}  \frac{    \left(  \widetilde{\chi}_0 \right)^{+-}_{ \boldsymbol{q} }\!(\omega ; T )  }{1 +  U
           \left(  \widetilde{\chi}_0 \right)^{+ -}_{ \boldsymbol{q} }\!\left(  \omega   ; T  \right)}  
  \nonumber \\
& =
 \frac{\overline{m} + \widetilde{\Lambda}_{\boldsymbol{q}}(\omega ; T)}{ \omega - \left[    -  U \widetilde{\Lambda}_{\boldsymbol{q}}(\omega  ; T)        \right]  }   .
  \label{Bethe-Salpeter q omega T adiabatic}
\end{align}
If $\widetilde{\Lambda}_{\boldsymbol{q}}(\omega ; T) $ is almost independent of $\omega$ at frequencies which are small with respect to the Stoner excitations, we can define the \emph{time-dependent magnon frequency} as
\begin{align}
\omega_{\boldsymbol{q}}(T) \equiv -  U \, \widetilde{\Lambda}_{\boldsymbol{q}}(0 ; T) . 
\end{align}

\section{The equilibrium case}
\label{app: equilibrium}

In equilibrium we have the exact identity
\begin{align}
\widetilde{A}(\omega, \Omega) = 2 \pi \delta(\Omega) A(\omega)
\label{equilibrium delta}
\end{align}
for the Fourier-Laplace transforms of the many-body functions of $\tau$ and $T$ involved in our derivation, since the latter do not depend on the total time $T$. This is a particular case of the adiabatic regime discussed in the main text, so all the equilibrium results can be immediately recovered from those valid in the adiabatic regime, by just removing the dependence of the exchange matrix (and, therefore, of the magnon frequencies) on the total time $T$. This can also be checked by using Eq.\eqref{equilibrium delta} to simplify Eqs.\eqref{bare susceptibility i j omega Omega} and \eqref{Bethe-Salpeter i j omega Omega}, and then by solving those equations directly, following exactly the same procedure which is discussed in the main text.

In particular, from Eq.\eqref{J EXCHANGE} we obtain the equilibrium exchange parameters as  
\begin{align}
  J_{i j}     =  &   \mathrm{i} \Sigma_{i S} \Sigma_{j S}  \lim_{\epsilon \rightarrow 0^+} \int_0^{\infty}  \mathrm{d} t \, \mathrm{e}^{- \epsilon t}  \Big[ 
\left( G^< \right)^{i \downarrow t}_{j \downarrow 0}    \left( G^> \right)_{i \uparrow t}^{j \uparrow 0}  \nonumber \\
&    - \left( G^> \right)^{i \downarrow t}_{j \downarrow 0}    \left( G^< \right)_{i \uparrow t}^{j \uparrow 0}      
   \Big] .
\end{align}
In the Matsubara representation, it is assumed that the statistical preparation of the initial state follows a thermal distribution. At zero temperature (or $\beta \rightarrow \infty$), the above expression is equivalent to
\begin{align}
  J_{i j}    & =   \Sigma_{i S} \Sigma_{j S}  \lim_{\beta \rightarrow \infty}  \int_{- \beta}^{\beta} \mathrm{d} \tau \, G^{i \downarrow}_{j \downarrow}(\tau) \,   G_{i \uparrow}^{j \uparrow}(-\tau)     ,
\end{align}
where $G(\tau)$ denotes a Matsubara Green function in the imaginary-time (here denoted as $\tau$) representation. Switching to the representation in terms of Matsubara frequencies $\omega_n$,
\begin{align}
G(\tau) = \frac{1}{\beta} \sum_{\omega_n} \mathrm{e}^{- \mathrm{i} \omega_n \tau} G(\mathrm{i} \omega_n) ,
\end{align}
as well as using Eq.\eqref{Stoner} in the equilibrium case, we obtain Eqs.\eqref{J Matsubara} of the main text.

\section{A useful sum rule for non-equilibrium Green functions}
\label{app: sum rule}

As mentioned in the main text, the fact that our theory is consistent with the Goldstone theorem, even out of equilibrium, can be immediately seen from the fact that 
\begin{align}
\sum_j \Lambda_{i j}(t, t') = 0.
\end{align}
The Goldstone theorem can also be checked in an alternative way by using a sum rule that we derive here, valid in and out of equilibrium within the THF approximation. The Dyson equations are given by Eqs.\eqref{Dyson THF SOH}, in particular
\begin{align}
&  - \mathrm{i} \partial_{t'} \! \left( G^{\lessgtr} \right)^{i \downarrow t}_{j \downarrow t'} \stackrel{\mathrm{THF}}{=} \! \left( G^{\lessgtr} \cdot T \right)^{i \downarrow t}_{j \downarrow t'}  + \! \left( G^{\lessgtr} \right)^{i \downarrow t}_{j \downarrow t'} \Sigma_{j \downarrow}(t') , \nonumber \\
&   \mathrm{i} \partial_{t'} \left( G^{\gtrless} \right)_{i \uparrow t}^{j \uparrow t'} \stackrel{\mathrm{THF}}{=} \left( T \cdot G^{\gtrless}  \right)_{i \uparrow t}^{j \uparrow t'}  + \Sigma_{j \uparrow}(t')  \left( G^{\gtrless} \right)_{i \uparrow t}^{j \uparrow t'}  .
\end{align} 
We multiply the first equation by $\left( G^{\gtrless} \right)_{i \uparrow t}^{j \uparrow t'}$ and sum over $j$; analogously, we multiply the second equation by $\left( G^{\lessgtr} \right)^{i \downarrow t}_{j \downarrow t'}$ and sum over $j$. We obtain
\begin{align}
&  \sum_j  \left( G^{\lessgtr} \right)^{i \downarrow t}_{j \downarrow t'} \left( - \mathrm{i} \overleftarrow{\partial}_{t'}  \right) \left( G^{\gtrless} \right)_{i \uparrow t}^{j \uparrow t'}   
\nonumber \\
& \stackrel{\mathrm{THF}}{=} 
   \left[  \left( G^{\lessgtr} \right)^{  \downarrow t}_{  \downarrow t'} \cdot  T(t') \cdot  \left( G^{\gtrless}  \right)_{  \uparrow t}^{  \uparrow t'} \right]^i_i  \nonumber \\
& \quad \quad  + \left[ \left( G^{\lessgtr} \right)^{  \downarrow t}_{  \downarrow t'} \cdot \Sigma_{  \downarrow}(t') \cdot  \left( G^{\gtrless} \right)_{  \uparrow t}^{  \uparrow t'}  \right]^i_i ,
   \label{sum rule 1st}
\end{align}
\begin{align}
& \sum_j   \left( G^{\lessgtr} \right)^{i \downarrow t}_{j \downarrow t'}  \left( \mathrm{i} \overrightarrow{\partial}_{t'} \right) \left( G^{\gtrless} \right)_{i \uparrow t}^{j \uparrow t'} \nonumber \\
&  \stackrel{\mathrm{THF}}{=} 
   \left[  \left( G^{\lessgtr} \right)^{  \downarrow t}_{  \downarrow t'} \cdot  T(t') \cdot  \left( G^{\gtrless}  \right)_{  \uparrow t}^{  \uparrow t'} \right]^i_i   \nonumber \\
& \quad \quad + \left[ \left( G^{\lessgtr} \right)^{  \downarrow t}_{  \downarrow t'} \cdot \Sigma_{  \uparrow}(t') \cdot  \left( G^{\gtrless} \right)_{  \uparrow t}^{  \uparrow t'}  \right]^i_i  .
   \label{sum rule 2nd}
\end{align} 
We subtract Eq.\eqref{sum rule 1st} from Eq.\eqref{sum rule 2nd}, divide by 2, and we obtain
\begin{align}
& \frac{1}{2} \, \mathrm{i} \overrightarrow{\partial}_{t'}  \left[   \left( G^{\lessgtr} \right)^{  \downarrow t}_{  \downarrow t'}    \left( G^{\gtrless} \right)_{  \uparrow t}^{  \uparrow t'} \right]^i_i  \nonumber \\
&  \stackrel{\mathrm{THF}}{=}   \left[ \left( G^{\lessgtr} \right)^{  \downarrow t}_{  \downarrow t'} \cdot \Sigma_{  S}(t') \cdot  \left( G^{\gtrless} \right)_{  \uparrow t}^{  \uparrow t'}  \right]^i_i  .
   \label{sum rule final}
\end{align} 
The sum rule Eq.\eqref{sum rule final} can be used to immediately check that Eqs.(44) and (45) indeed satisfy
\begin{align}
\sum_j \left[ J_{ij}(t, t') + X_{i j}(t, t') \right] = 0 ,
\end{align} 
which is in agreement with the Goldstone theorem.

\end{document}